\documentclass[aps,prx,twocolumn,superscriptaddress,longbibliography]{revtex4-1}

\usepackage{amsmath,amsfonts,amssymb}
\usepackage{braket}
\usepackage{graphicx}
\usepackage{tikz}
\usetikzlibrary{shapes,arrows,positioning,calc}
\usepackage{textcomp}
\usepackage[version=4]{mhchem}
\usepackage[hidelinks]{hyperref}
\usepackage{enumitem}

\newcommand{\mb}[1]{{\mathbf{#1}}}

\newcommand{\ul}[1]{\underline{#1}}

\newcommand{\be}{\begin{equation}}
\newcommand{\ee}{\end{equation}}

\begin{document}

\title{Searches for massive neutrinos with mechanical quantum sensors}

\author{Daniel Carney}
\affiliation{Physics Division, Lawrence Berkeley National Laboratory, Berkeley, CA}
\author{Kyle G. Leach}
\affiliation{Department of Physics, Colorado School of Mines, Golden, CO}
\affiliation{Facility for Rare Isotope Beams, Michigan State University, East Lansing, MI}
\author{David C. Moore}
\affiliation{Wright Laboratory, Department of Physics, Yale University, New Haven, CT}
\date{\today}

\begin{abstract}
The development of quantum optomechanics now allows mechanical sensors with femtogram masses to be controlled and measured in the quantum regime. If the mechanical element contains isotopes that undergo nuclear decay, measuring the recoil of the sensor following the decay allows reconstruction of the total momentum of all emitted particles, including any neutral particles that may escape detection in traditional detectors. As an example, for weak nuclear decays the momentum of the emitted neutrino can be reconstructed on an event-by-event basis. We present the concept that a single nanometer-scale, optically levitated sensor operated with sensitivity near the standard quantum limit can search for heavy sterile neutrinos in the keV--MeV mass range with sensitivity significantly beyond existing laboratory constraints. We also comment on the possibility that mechanical sensors operated well into the quantum regime might ultimately reach the sensitivities required to provide an absolute measurement of the mass of the light neutrino states. 
\end{abstract}

\maketitle

\section{Introduction}
While the Standard Model (SM) of particle physics provides an extremely successful description of the known fundamental particles and their interactions, astrophysical and laboratory measurements indicate that it is incomplete. In particular, it does not account for the origin of the small, but non-zero, neutrino masses observed in neutrino oscillation experiments~\cite{Workman:2022}. Although minimal versions of the SM contain only left-handed, massless neutrinos, extensions incorporating non-zero masses typically require right-handed neutrinos~\cite{deGouvea_nu_mass_review}. Such ``sterile'' neutrinos would not directly participate in weak interactions~\cite{ALEPH:2005ab_lep_z}, but may mix with the active neutrinos to produce observable effects~\cite{Dasgupta:2021ies_sterile_review}. Searches for sterile neutrinos over a wide range of mass scales is a current focus of the neutrino community~\cite{Dasgupta:2021ies_sterile_review,Giunti:2019aiy_eV_steriles}. It is also possible that $\gtrsim$ keV sterile neutrinos could constitute some (or all) of the relic dark matter density~\cite{Boyarsky:2018tvu_sterile_dm_review}. 

Most searches for sterile neutrinos aim to detect oscillations between the SM neutrinos produced in weak interactions and possible sterile states~\cite{Giunti:2019aiy_eV_steriles}. Due to constraints on the neutrino energy and size of the detectors in such experiments, these searches have the greatest sensitivity to neutrinos with masses $\lesssim~$10s of eV. For higher mass neutrinos, searches for distortions in the spectrum of electrons emitted by nuclear $\beta$ decays provide among the most stringent laboratory constraints in the $\sim$keV--MeV mass range~\cite{HOLZSCHUH1999247_ni63,HOLZSCHUH20001_s35,SCHRECKENBACH1983265_cu64, DEUTSCH1990149_f20, PhysRevLett.86.1978_re187, Kraus:2012he_mainz,Belesev:2013cba_troitsk,2018JETPL.108..499D_ce144pr144, PhysRevLett.126.021803_beest}. If sterile neutrinos constitute a significant fraction of dark matter, limits on X-ray emission from their radiative decays may also provide constraints in this mass range~\cite{Dasgupta:2021ies_sterile_review, PhysRevD.75.053005}. Accelerator based searches for heavy neutral leptons provide the dominant constraints at $\sim$MeV--GeV masses~\cite{PhysRevD.100.073011_pienu}.
 
A complementary approach is to fully reconstruct the kinematics of all particles (except the neutrino) in the final state of a weak nuclear decay~\cite{SHROCK1980159_kinem,PhysRevD.46.R888_kinem,PhysRevLett.90.012501_trap,PhysRevC.58.2512_ar37,2021QS&T....6b4008M_hunter_qst, 2019NJPh...21e3022S_hunter_smith}. This approach avoids the need to detect small distortions in the $\beta$ spectrum, which becomes increasingly challenging for small branching ratios to decays emitting a keV--MeV neutrino. 
Here we describe a new approach to kinematic reconstruction of weak decays using mechanical quantum sensors. This proposal takes advantage of the rapid development of levitated optomechanics for nanoscale objects over the past decade~\cite{Levitodynamics2021,2020RPPh...83b6401M_millen_review, Fadeev:2020xzw_ferro_qst}. These techniques now allow objects with diameters $\sim$100~nm to be trapped in high vacuum, and their motional state cooled to the ground state of the trapping potential~\cite{Delic2020,Tebbenjohanns2021,magrini2021real,Ranfagni:2021pnb_ground_state,Kamba:2022cgy_ground_state}. Operating such systems in the quantum regime has recently enabled demonstrations of cavity-free squeezing of the optical fields interacting with the particles~\cite{ Militaru:2022pen_opt_squeeze, Magrini:2022rlc_opt_squeeze}. Work is ongoing to prepare non-classical motional states and to read out the position or momentum of such objects beyond the ``standard quantum limit'' (SQL)~\cite{Levitodynamics2021,PhysRevLett.128.143601_oriol_squeezing,PhysRevResearch.2.013052_filip_squeeze,PhysRevLett.127.023601_deloc, Neumeier:2022czd_squeeze, Zhou:2022fr_large_super}. 

Applications of levitated optomechanical systems to tests of fundamental physics include searches for tiny forces or momentum transfers from, e.g., scattering of dark matter particles~\cite{Monteiro2020DM, Afek:2021vjy_coherent, CarneyMechanical, Moore_2021, Riedel2013, Riedel2017}, modifications to gravity~\cite{PhysRevD.104.L061101_gratta_grav, PhysRevD.104.L101101_ulbricht_grav}, or the presence of fractionally charged dark matter particles~\cite{AfekMCP2021, Priel:2021wxa_charge, Moore2014}. Ambitious future proposals with such systems aim to test the foundations of quantum mechanics with massive particles~\cite{Carlesso:2022pqr, BELENCHIA20221_space, PhysRevResearch.2.013057_csl, PhysRevResearch.2.043229_csl}, detect high frequency gravitational waves~\cite{PhysRevLett.128.111101_geraci_gw, 2022arXiv220410843W_geraci_hex_trap}, or even entangle microparticles only through their mutual gravitational attraction~\cite{Bose2017, Marletto2017,Carney:2018ofe}. Extending such techniques to gram-scale objects has also been recently proposed for detection of coherent scattering of reactor neutrinos~\cite{Kilian:2022kgm_bose_nu}. As described here, these systems can also be used to reconstruct the mass of the emitted neutrinos in weak nuclear decay through precision measurements of the recoils from embedded radioisotopes in nanoscale particles.

Similar techniques as those proposed here were recently used to set world-leading constraints on sub-MeV mass sterile neutrinos by implanting $^7$Be in energy resolving superconducting sensors~\cite{PhysRevLett.126.021803_beest}. However, the mechanical sensors proposed here provide important complementarity to energy resolving measurements of the nuclear recoil. Since the momentum carried by secondary atomic particles (i.e. Auger $e^-$ or X-rays) emitted following the decay is much smaller than the nuclear recoil momentum, momentum resolving measurements largely avoid smearing from the interactions of different parent atoms with their surroundings~\cite{Samanta:2022siw}. In contrast, these smearing effects are much more significant in measurements of the nuclear recoil energy. If all decay products are stopped within the nanoparticle except the $\nu$ (see Sec.~\ref{sec:steriles}--\ref{sec:light_nu}), momentum conservation guarantees that the center-of-mass (COM) of the particle will carry momentum exactly equal and opposite to the neutrino, regardless of any internal excitations or atomic broadening.  

Compared to other existing and proposed techniques to probe weak neutrino admixtures~\cite{Acero:2022wqg}, including those using trapped atoms~\cite{PhysRevD.75.053005,DORNER200095_coltrims,2021QS&T....6b4008M_hunter_qst, 2019NJPh...21e3022S_hunter_smith}, the techniques proposed here benefit from the high density of the solid source material allowing a large number of decaying nuclei to be confined. In addition, essentially any isotope of interest can be confined in the solid particle, with no requirement that the species of interest be compatible with direct laser trapping and cooling. The recoil measurement is performed by collecting light scattered by the particle, which can be used to infer its position as a function of time (and thus reconstruct a change in the particle momentum). This measurement depends only on the scattering of light by a sub-wavelength particle, and does not strongly depend on the detailed particle properties or readout laser wavelength. These advantages may permit relative electron-flavour admixtures to keV--MeV mass states of $\lesssim 10^{-8}$ or lower to be probed with a modest array of nanoparticles in a month-long exposure (see Sec.~\ref{sec:proj_sens_steriles}), which is more than 4 orders of magnitude beyond existing laboratory constraints in this mass range.

Finally, the ultimate sensitivity reached by mechanical measurements with nanoscale particles might someday permit even the light neutrino masses to be detected (see Sec.~\ref{sec:light_nu}), potentially employing ``ultra-low $Q$-value'' decays, if such transitions could be identified (here, $Q$ refers to the total energy released in the decay). If so, the ability to prepare the COM motion of the nanoparticle in a sufficiently narrow momentum state ($\lesssim$200 meV) permits the neutrino to be in a spatially delocalized state, which is necessary for a precise determination of the neutrino momentum but is typically not possible for a solid source where the decay location itself is localized to atomic distances~\cite{PTOLEMY:2022ldz_theory,Cheipesh:2021fmg}. Controlling the charge state of the nanosphere could also be leveraged to trap any emitted low energy electrons, allowing a fully two-body decay for the low energy decays of interest for light neutrinos.

The paper is structured as follows. In Sec.~\ref{sec:mass_measurements}, we describe the general principles for mass reconstruction with mechanical nanoscale sensors. We show in Sec.~\ref{sec:steriles} that searches for sterile neutrinos in the keV--MeV mass range with orders-of-magnitude higher sensitivity than existing experiments can be performed, using existing technologies. In Sec.~\ref{sec:light_nu}, we comment on the possibility that the unique properties of levitated optomechanical systems---operated well into the quantum regime---might someday reach sensitivity to the masses of the much lighter SM neutrinos. Appendix \ref{manybody} provides a simple three-body model for the decay event inside a nanosphere.

\section{Weak nuclear decay inside mechanical sensors}
\label{sec:mass_measurements}

In a weak nuclear decay, an electron neutrino, $\nu_e$ (or its antiparticle, $\bar{\nu}_e$), is produced at the interaction vertex. These flavor states are superpositions 
\be
\ket{\nu_e} = \sum_i U_{ei}^* \ket{\nu_i}
\ee
of the neutrino mass eigenstates, labeled by $i \in \{ 1, 2, 3, \ldots \}$. The mass eigenstates satisfy mass-energy relations of the form $E_i^2 = m_i^2 + \mb{p}^2$, while there is simply no analogous relation for the flavor eigenstates.  For all flavor eigenstates $\ket{\nu_{\alpha}}$, $\alpha \in \{ e, \mu, \tau \}$, the matrix $U_{\alpha i}$ rotating between these bases is called the Pontecorvo-Maki-Nakagawa-Sakata (PMNS) matrix~\cite{1968JETP...26..984P,Maki:1962mu}.

In the SM, the PMNS matrix is a $3\times3$ unitary transformation acting on three mass eigenstates. In Sec.~\ref{sec:steriles}, we will primarily be concerned with the possibility that there is one additional state (or possibly multiple such states) with a mass eigenvalue $m_4 \gtrsim 1~{\rm keV}$, which is much larger than the SM neutrino masses, $\sum_{i=1}^3 m_i \lesssim 0.2~{\rm eV}$~\cite{DES:2021wwk_nu_mass}. In this case, the PMNS matrix is assumed to be extended to a $4\times4$ unitary matrix, including mixing with this heavier state. A number of laboratory bounds already constrain $|U_{e4}|^2 \lesssim 10^{-3}$ for keV--MeV mass sterile $\nu$, along with astrophysical and cosmological bounds that can be much more stringent but are typically model-dependent~\cite{Dasgupta:2021ies_sterile_review, Hagstotz:2020ukm}. Thus we are looking for relatively rare events---in a given weak decay, the most likely final states are any of the light SM neutrinos, with possibly a small branching ratio to the heavier sterile states.

\begin{figure*}[t]
  \centering
  \includegraphics[width=\textwidth]{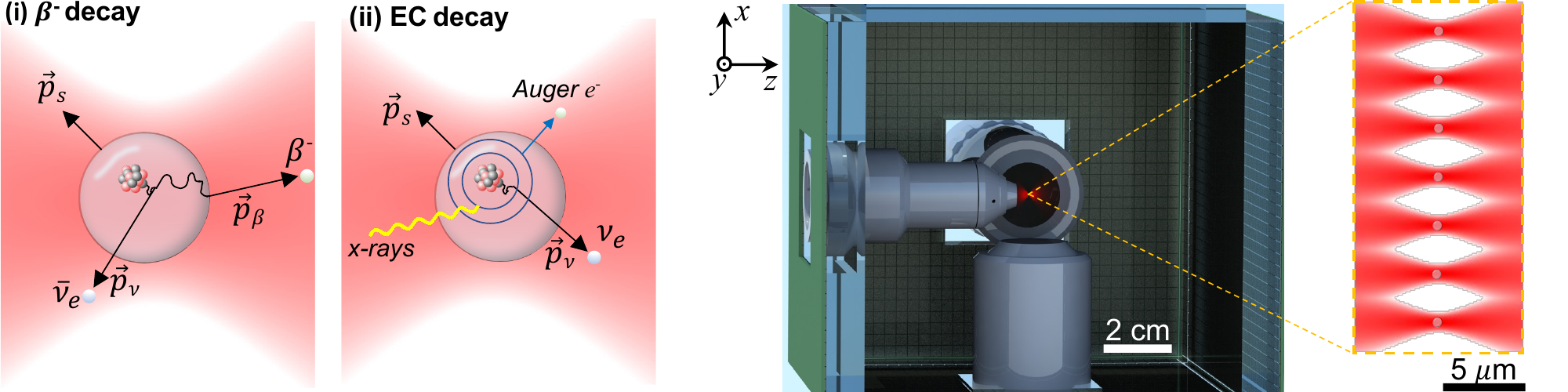}
  \caption{Schematic of detection concept described in the text. (Left) Zoom to a single spherical nanoparticle, schematically showing (i) a $\beta^-$ decay, for which the nuclear recoil is stopped in the nanosphere and the $e^-$ and $\bar{\nu}_e$ exit the particle; and (ii) an electron capture (EC) decay, for which only the $\nu_e$, and lower energy atomic X-rays or Auger $e^-$ are emitted, such that the nanosphere recoils with momentum nearly equal and opposite to that of the emitted $\nu_e$. (Right) Cut-away view of the trap concept and surrounding detectors. The trap is formed by a low profile, high numerical aperture objective, with two additional long working distance objectives positioned perpendicularly to the trapping objective to collect scattered light. Scintillator panels instrumented with SiPM arrays surround the trap on all sides to reconstruct emitted secondary particles. (Right inset) Zoom in on the trapping region, schematically showing traps for multiple particles formed with a tweezer array~\cite{Roichman2007,Lester2015,Endres2016,Barredo2018,Covey2019,PhysRevX.12.021027,Arita:18,Rieser:2022nye}. }
  \label{fig:concept}
\end{figure*}

The general decay processes of interest here produce an electron flavor neutrino, starting from parent particles $A$ and producing additional daughter particles $B$, $A \to B \nu_e$. The examples we will focus on are nuclear $\beta$-decay of neutrons $n \to p e \bar{\nu}_{e}$ and nuclear electron capture (EC), $e p \to n \nu_{e}$. A parent particle in an initial momentum eigenstate $\ket{\psi}_{\rm in} = \ket{\mb{p}_A}$ will decay into a superposition of the form
\begin{align}
\begin{split}
\label{AtoBnu}
\ket{\psi}_{\rm out} & = \sum_i \int d^3\mb{p}_i d^3\mb{p}_{B,i} \delta^4(p_A - p_i - p_{B,i}) \\
& \mathcal{M}_{e}(\mb{p}_A;\mb{p}_i,\mb{p}_{B,i}) U^*_{e i}  \ket{\mathbf{p}_i} \otimes \ket{\mathbf{p}_{B,i}},
\end{split}
\end{align}
where the first ket labels the various decay products $B$, the second labels the neutrino, and $\mathcal{M}$ is the probability amplitude for the decay $A \to B \nu_{e}$. Notice that in \eqref{AtoBnu}, the mass eigenstates appear in a coherent superposition, entangled with the $B$ particles through total energy-momentum conservation~\cite{Cohen:2008qb}. On each branch of the wavefunction, total four-momentum conservation reads $p_A^{\mu} = p_{i}^{\mu} + p_{B,i}^{\mu}$, which implies
\be
\label{mi2}
m_i^2 = -(p_A^{\mu} - p_{B,i}^{\mu})^2.
\ee
Thus if we can resolve the momentum difference $(p_A - p_B)^2$ sufficiently well, we can determine the mass eigenvalues of the neutrinos. The probability of the final state neutrino being in a given mass eigenvalue $i$ is proportional to $|U_{e i}|^2$ as well as a phase space factor.

Our proposed experimental strategy is to embed specific unstable radioisotopes into a mesoscopic (sub-micron) mass, whose COM degree of freedom (DOF) is continuously monitored for changes in momentum (see Fig. \ref{fig:concept}). Consider a nanosphere of some solid material of mass $m_s$, implanted with an unstable isotope $A$ of atomic mass $m_A$. After the decay $A \to B \nu_{e}$, the end state $B$ will include a daughter isotope, which will be stopped within the solid particle (see below), as well as potentially other decay products $B'$ which may either be stopped or exit the nanosphere. Measurement of the momentum kick to the nanosphere, as well as the momentum of any secondary particles that exit the nanosphere, can be used to kinematically reconstruct the neutrino momentum. While similar proposals for kinematic reconstruction of weak nuclear decays have been studied previously~\cite{SHROCK1980159_kinem,PhysRevD.46.R888_kinem,PhysRevLett.90.012501_trap,PhysRevC.58.2512_ar37,2021QS&T....6b4008M_hunter_qst, 2019NJPh...21e3022S_hunter_smith}, such proposals have not considered the case that the decaying radioisotope is embedded in a nanoscale solid particle.

The final state after the decay is a complex joint state involving the outgoing neutrino and nanosphere COM, as well as any additional decay products and the various internal excitations and rotational modes of the nanosphere. However, crucial to our proposal is that overall energy-momentum conservation applies in the form
\be
\label{sphereconservation}
\begin{pmatrix} E_s \\ \mb{p}_s \end{pmatrix} = \begin{pmatrix} E_{s'} \\ \mb{p}_{s'} \end{pmatrix} + \begin{pmatrix} E_i \\ \mb{p}_i \end{pmatrix} + \begin{pmatrix} E_{B',i} \\ \mb{p}_{B',i}  \end{pmatrix}.
\ee
Here, $B'$ represents the subset of decay products which exit the nanosphere, $E_s$ and $E_{s'}$ are the initial and final energy of the nanosphere, and $\mb{p}_s$ and $\mb{p}_{s'}$ are its initial and final momenta, with change in momentum $\Delta \mb{p}_s = \mb{p}_{s'}-\mb{p}_s$. From measurement of the final-state momenta and energies, one could in principle rearrange \eqref{sphereconservation} to give a direct neutrino mass-squared reconstruction analogous to \eqref{mi2}~\cite{2019NJPh...21e3022S_hunter_smith}. However, the internal energy of the nanosphere may be difficult to measure in practice, and for the decays of interest in Sec.~\ref{sec:steriles} measurement of these energies is not required to reach the desired sensitivity. 

Instead, sensitivity to the keV--MeV scale mass states considered here can be reached by considering only the portion of \eqref{sphereconservation} related to three-momentum conservation. Measurements of the nanosphere recoil momentum, and the angle and energy of emitted secondaries permit reconstruction of the neutrino momentum vector $\mb{p}_i = -(\Delta \mb{p}_s + \mb{p}_{B',i})$. The value of the neutrino mass $m_i$ will distort the probability distribution of final-state momenta $P(\mb{p}_i)$ relative to the case where the neutrinos are massless, because some of the decay energy is required to produce the neutrino rest mass, reducing its momentum. This effect is most pronounced for non-relativistic neutrinos, i.e. those whose rest mass is an appreciable fraction of their total energy. The details differ somewhat in a $1 \to 2$ decay (electron capture) and a $1 \to 3$ decay ($\beta$-decay), and are discussed separately in Secs.~\ref{sec:ec_decays}--\ref{sec:beta_decays}. See Figs. \ref{fig:recoil_spec_EC} and \ref{fig:recoil_spec_beta}
for examples.

In either case, for the $Q\sim {\rm MeV}$ decays considered here, the recoiling nucleus will carry $\lesssim 100~{\rm eV}$ of energy due to its large mass. Extrapolating the expected range of such recoils from $>$keV energies~\cite{lindhard1963range} to the extremely low energies considered here indicates that the recoiling nucleus would typically be expected to stop within a $10-100~{\rm nm}$ nanosphere (see further discussion in Sec.~\ref{sec:sterile_backgrounds} below). On the other hand, the charged or radiative secondaries will often be ejected. From an experimental standpoint, these secondaries can provide a convenient trigger: i.e., we can study momentum kicks to the nanosphere conditioned on knowing a decay happened by detection of a secondary particle. As we describe in detail below, the EC secondaries have sufficiently low momentum that we will not need to measure them precisely, whereas in the $\beta^-$ case we will need reasonably accurate measurements of the energy and emission angle.

We will consider using mechanical sensors of mass $m_s$ in harmonic traps of frequency $\omega_s$, read out with a sensitivity around the ``standard quantum limit'' (SQL) \cite{caves1981quantum}, which describes the accuracy of simultaneous weak measurements of position and momentum when the dominant errors arise from the measurement process itself, due to the Heisenberg uncertainty relation~\cite{RevModPhys.82.1155}. A measurement at the SQL can determine $\Delta \mb{p}_s$ (for an impulse delivered over time $\tau \ll \omega_s^{-1}$) with accuracy~\cite{PhysRevB.70.245306, Ghosh:2019rsc}
\be
\label{sql}
\Delta p_{\rm SQL} = \sqrt{\hbar m_s \omega_s} = 15~{\rm keV} \times \left( \frac{m_s}{1~{\rm fg}} \right)^{1/2} \left( \frac{\omega_s/2\pi}{100~{\rm kHz}} \right)^{1/2}.
\ee
Examples of such devices in current operation include \cite{Tebbenjohanns2021, magrini2021real, Kamba:2022cgy_ground_state}. These numbers suggest that searches for heavy sterile neutrinos with $m_4 \gtrsim 100~{\rm keV}$ and with branching ratios $|U_{e4}|^2 \lesssim 10^{-4}$ will be relatively straightforward with existing technologies, using trapped nanospheres with radius $\sim 100~{\rm nm}$. On the other hand, they also suggest that direct mass reconstruction of the light neutrino masses will require substantially smaller sensor masses, potentially down to single ions, or more massive systems operated well below the SQL (see Sec. \ref{sec:light_nu}).

\section{Detector and search design}
\label{sec:steriles}
The concept for sterile neutrino searches is shown in Fig.~\ref{fig:concept}, along with a possible experimental design. To reconstruct the nanosphere momentum, the position of the nanosphere is measured continuously in each of the three independent Cartesian degrees-of-freedom of the COM motion using the light scattered by the nanosphere, and its momentum is reconstructed from these measurements. As described in Sec.~\ref{sec:mass_measurements}, quantum limited measurement of the nanosphere position has been reached in existing systems~\cite{Tebbenjohanns2021, magrini2021real, Kamba:2022cgy_ground_state}. Avoiding thermal noise requires operation at ultra-high vacuum (UHV) pressure $\lesssim 10^{-8}$~mbar and high numerical aperture (NA) collection optics surrounding the trap to collect as much of the information carried in the light scattered by the nanosphere as possible~\cite{Tebbenjohanns2019Optimal, Tebbenjohanns2021, magrini2021real, Kamba:2022cgy_ground_state} . 

In the $z$ DOF, the backscattered light contains the majority of information about the $z$ displacement, and the trapping objective itself can be used to collect this light~\cite{Tebbenjohanns2019Optimal, magrini2021real}. For the $x$ and $y$ degrees-of-freedom, high light collection competes with the requirement to detect coincident secondary particles from the decay. For this reason, high NA, long working distance objectives are placed on only two sides of the trapping region to collect side scattered light, while the remaining three sides of the trap have no intervening material between the nanosphere and scintillator slabs positioned in a cubic arrangement at $\sim$5~cm distance on each side from the trap location~\footnote{More complex designs are possible. E.g., the surfaces of the scintillators could be coated with a reflective coating and act as mirrors to reflect the imaging light back onto the particle and into the opposite objective~\cite{Dania:2021dgp_self_interfere}, while not affecting detection of the secondary particles}. The slabs are instrumented with an array of pixellated light detectors (e.g., silicon photomultipliers [SiPMs]) to detect the energy and emission angle of the $e^-$ and X-rays emitted by the decays. Additional panels are positioned around the objectives, to provide further vetoing of backscattered $e^-$ or other backgrounds.  The SiPM dark noise in the $\sim$10~$\mu$s detection time for the recoil is sub-dominant for few mm$^2$ pixels, allowing the entire detector including the trap, vacuum chamber, and SiPM array to be operated at room temperature.

\begin{table*}[t]
    \centering
    \begin{tabular}{c | c c c c c c}
Isotope &	$Q_{EC}$ [keV]	& $T_{1/2}$ [day] &	Auger energies [keV] &	X-ray energies [keV] &	$\gamma$ energies [keV] &	Live time [sphere  days]\\
\hline
$^7$Be &	862 & 53.2 &	0.045 (100\%) &	&	477 (10.6\%) &  24\\
$^{37}$Ar &	814 & 35 &	2.4 (81\%) &	2.6 (8.2\%), 2.8 (0.5\%) &	& 9 \\
$^{49}$V &	602 & 330 &	4.0 (69\%), 0.4 (149\%) &	4.5 (17.1\%), 4.9 (1.9\%) & &	119 \\
$^{51}$Cr &	752	& 27.7 &	4.4 (66\%), 0.5 (146\%) &	4.9 (19.4\%), 5.4 (2.2\%)	& 320.1 (9.9\%) &	9\\
$^{68}$Ge &	107	& 271 &	8.0 (42\%), 1.1 (122\%)	& 9.2 (39\%), 10.3 (5\%) &	&	147 \\
$^{72}$Se &	361	& 8.4 &	9.1 (53\%), 1.2 (167\%) & 10.5 (62\%), 11.7 (8\%) &	45.9 (57\%) & 4 \\
    \end{tabular}
    \caption{Selected EC isotopes of interest for searches for heavy neutrino mass searches between $\sim$0.1-1~MeV. The $Q_{EC}$ value for the decay and half-life are listed for each candidate, along with the energies and intensities (i.e., percentage of primary decays of the parent nucleus for which a given secondary particle is emitted) of the secondary Auger $e^-$, X-rays, or for $\gamma$s from decays to nuclear excited states. The live time required to observe $10^4$ events (in number of nanospheres $\times$ days) assuming 100~nm diameter SiO$_2$ nanospheres with 1\% loading by mass for each isotope, 40\% detection efficiency for the secondary particles, and the detector parameters described in the text is given in the last column. Isotope data are taken from~\cite{ENSDF}.}
    \label{tab:ec_isotopes}
\end{table*}

To estimate the sensitivity to $m_4$, a Monte Carlo (MC) simulation of the decay and detection process was performed. This simulation produces the primary particles (i.e., the $\nu$ and recoiling nucleus) as well as secondary particles ($\gamma$s, $\beta$s, X-rays, and Auger $e^-$) for each decay based on the decay products and branching ratios from~\cite{ENSDF}. For $\beta^-$ decays, the $e^-$ energy is sampled assuming the spectral shape for an allowed transition~\cite{Morita1963PThPS..26....1M}. However, an advantage of the proposed technique is that it does not depend on the detailed spectral shape for a given decay, or precise knowledge of that shape. The simulation assumes the nuclear recoil fully stops in the particle (transferring its full momentum to the COM), while all $\gamma$ and X-ray photons leave the particle with no energy loss. Energy loss for electrons from $\beta$ decay or Auger $e^-$ within the particle is randomly simulated using calculations of the electronic stopping power from~\cite{ASHLEY1981127_sio2_stopping,1992esta.rept.....B}, assuming a decay location uniformly distributed in the nanosphere volume. Noise is added to the nanosphere momentum reconstruction assuming an information collection efficiency in each direction of $(\eta_x, \eta_y, \eta_z) = (0.4, 0.4, 0.6)$~\cite{Tebbenjohanns2019Optimal,Tebbenjohanns2021,magrini2021real}. The angle of the emitted secondary is assumed to be reconstructed with an accuracy of 0.02~rad. This corresponds, e.g., to 1~mm resolution on the detection of the secondary particle interaction location at a distance of 5~cm from the nanosphere, based on the expected few mm pixellization of the SiPM array. The fractional energy resolution for detection of emitted MeV scale $e^-$ is assumed to be $\sigma_E/E = 0.01\sqrt{(\mathrm{1\, MeV})/E}$, consistent with a light collection efficiency for the SiPM array $\gtrsim 25$\% (or, e.g., with pixellized Si particle detectors which can provide improved energy resolution at lower energies). For EC decays, the energy of the secondary Auger $e^-$ and X-rays does not need to be precisely determined, and their emission angle can be reconstructed with the SiPM array even with relatively poor energy resolution. We assume that the secondary particle can be determined to be either a $\gamma$ or $e^-$ with high accuracy by monitoring the change in the net electric charge of the nanosphere before and after the decay~\cite{Moore2014,Frimmer2017}.

As described above, the trapping and measurement techniques proposed here allow a variety of isotopes to be considered as the source of the decay. The primary requirement is that the nanospheres have sufficiently low optical absorption at the trapping laser wavelength that internal heating of the particle does not cause melting or particle loss in vacuum~\cite{Monteiro2017,Moore_2021}. This enables consideration of doping (or implanting) nearly any isotope of interest in a silica nanosphere at percent-level concentrations or higher. 

Biomedical applications have produced spherical nano- and micro-particles including, e.g., P~\cite{hoss2015}, S~\cite{C3CC41062E}, Y~\cite{WESTCOTT2016351} doped into SiO$_2$. Gaseous elements, including H$_2$ have been stably stored at $>$percent level concentrations in SiO$_2$ microparticles (including hollow shells) by diffusion into the solid particle at high temperature and pressure~\cite{DURET1994757}. It is also possible to produce nanoparticles from a specific material containing a given isotope, provided it has sufficiently low optical absorption. E.g., yttria and vanadium oxide are high quality optical materials at infrared wavelengths and can be fabricated into nanoparticles~\cite{Dhanaraj200111098, cueff2020vo2}. Polymers are also commonly used for producing optically transparent nanoparticles, and, e.g., $^3$H could be embedded at high concentrations in tritiated-polymer~\cite{HU2017171} nanospheres. While a detailed study of the optimal fabrication method of particles containing a specific isotope is not considered here, it is expected that there are possible fabrication techniques reaching 1--10\% or higher doping levels for many or all of the isotopes considered.

\subsection{Electron capture decays}
\label{sec:ec_decays}
For unstable neutron deficient isotopes with $Q_{EC}<2m_e$, $\beta^+$ decay is energetically forbidden and thus only EC decay is possible, in which a nucleus X with atomic number Z and mass number A decays to the daughter Y through capture of an atomic electron and emission of a $\nu_e$: 
\begin{equation}
\label{ec}
  \ce{^{A}_{Z}X} + e^- \rightarrow \ce{^{A}_{Z-1}Y} + \nu_e.
\end{equation}
This provides a two-body final state with an emitted $\nu_e$, for which a reconstruction of the recoiling daughter nucleus Y can provide a measurement of the $\nu_e$ momentum using the techniques discussed above. At $\lesssim$ps time scales following the decay, the atomic vacancy left by EC will be filled by a higher-lying electron, typically producing additional Auger $e^-$ or X-rays with total energy of several keV for $A < 100$ isotopes. For 100~nm scale particles, $\gtrsim$keV secondary $e^-$ or X-rays will essentially always leave the nanosphere and provide a trigger for the event. However, the momentum carried by such secondaries is $\lesssim$100~keV even for the Auger $e^-$ (and only $\sim$keV for the X-rays), while the $\nu$ can carry momentum of several hundred keV for the decays of interest. Thus, such secondaries provide a subdominant contribution to the total momentum imparted to the nanosphere, although accurate reconstruction of the momentum carried by Auger $e^-$ can be important for $m_4 \lesssim 100$~keV. In particular, any attempt to determine the light SM neutrino masses in this way would require very accurate measurements of these Auger $e^-$, or stopping the $e^-$ within the particle itself (see Sec.~\ref{sec:light_nu}).

\begin{figure}[t]
  \centering
  \includegraphics[width=\columnwidth]{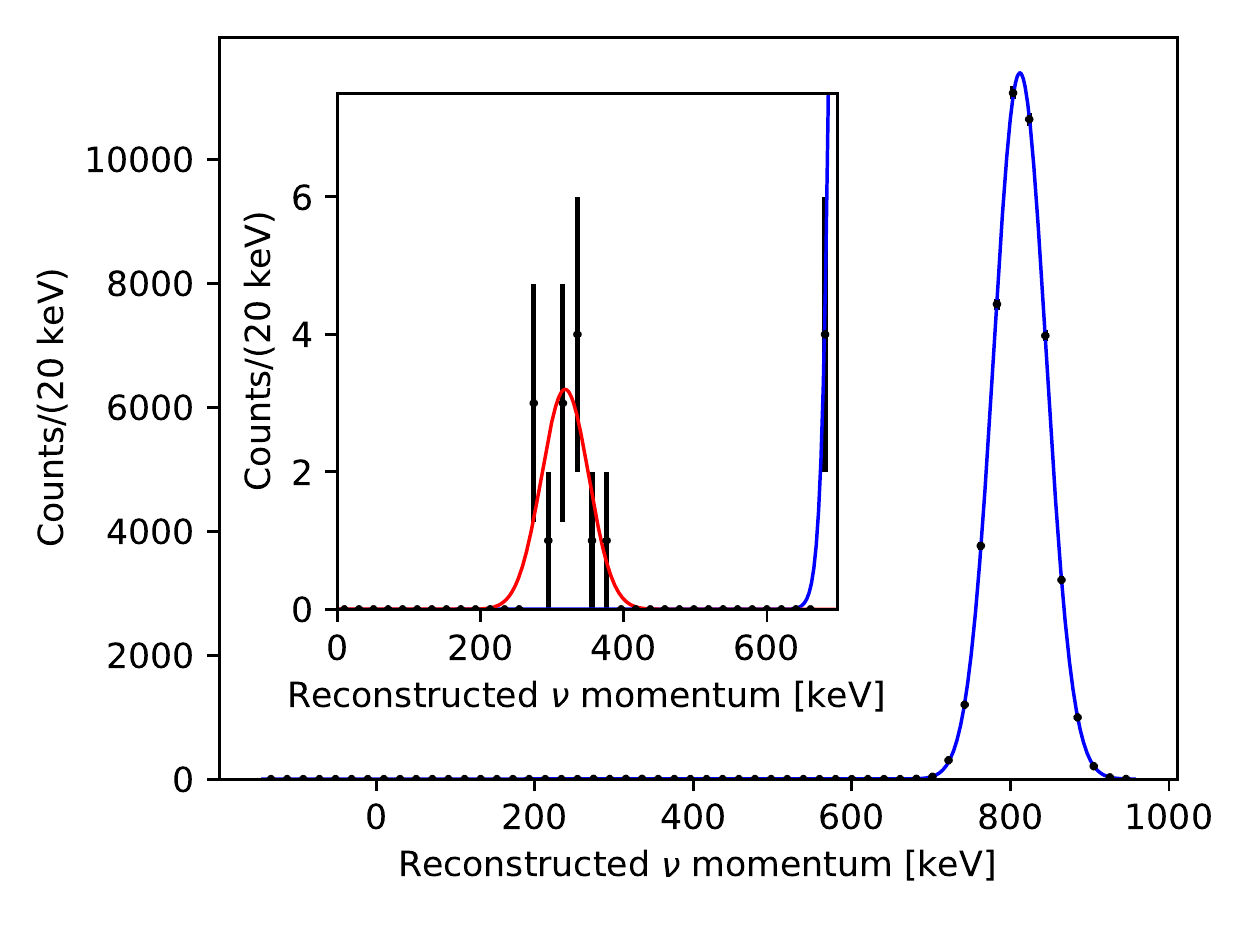}
  \caption{Simulated spectrum for the reconstructed neutrino momentum, $|\mb{p}_\nu|$, for a 100~nm diameter nanosphere containing 1\% by mass of the EC-unstable isotope $^{37}$Ar. We assume an exposure of 30~days and 40\% collection efficiency for the emitted Auger $e^-$ or X-ray. Also overlaid (and visible in the inset) is the expected spectrum arising for a heavier sterile state with $m_4 = 750$~keV and at the existing upper limits for a sterile $\nu$ at this mass, $|U_{e4}|^2 = 2\times10^{-4}$~\cite{PhysRevLett.126.021803_beest}. Backgrounds and misreconstructed events are not included in the simulation (see Sec.~\ref{sec:sterile_backgrounds} for a detailed discussion of the required background rejection to meet this assumption). The black data points with error bars indicate the simulated counts for a typical experimental realization, along with the best fit to these data for decays to SM neutrinos (blue) and the heavier sterile state (red).}
  \label{fig:recoil_spec_EC}
\end{figure}

For sufficiently high neutrino mass that the secondary $e^-$ or X-rays can be neglected, the difference in momentum between a decay to a sterile state with mass $m_4$ and a light neutrino with $m_i \approx 0$ (for $i=1,2,3$) is 
\begin{equation}
|\mb{p}_4| - |\mb{p}_{i}| \approx Q\left(\sqrt{1 - (m_4/Q)^2} - 1\right).
\label{eq:momentum_diff}
\end{equation}
In particular, as $m_4$ approaches $Q$, the presence of a heavy sterile state can be identified by a small peak separated in reconstructed momentum from the decays to the approximately massless state, as shown in Fig.~\ref{fig:recoil_spec_EC}. 

Table~\ref{tab:ec_isotopes} lists the EC candidate isotopes considered here, which have half-lives $\lesssim$1~year to provide a sufficient number of decays in a nanoscale particle. Of these, an especially simple case is $^{37}$Ar ($T_{1/2}=35$~d) which has a pure transition to the nuclear ground state of $^{37}$Cl with $Q_{EC} = 814$~keV. The EC decay of $^{37}$Ar is typically accompanied by a 2--3~keV Auger $e^-$ or X-ray, that can be used in coincidence with detection of the nanosphere recoil to identify decay events and avoid backgrounds. In this case, the nanosphere recoils with a momentum that is nearly equal and opposite to the momentum of the outgoing $\nu$, and the measurement accuracy of the Auger $e^-$ or X-ray energy and angle provides a sub-dominant correction to the reconstruction of $m_4$. 

Figure~\ref{fig:recoil_spec_EC} shows a MC simulation of the nanosphere recoil momentum spectrum for $10^5$ $^{37}$Ar decays, with the expected spectrum for an example of a heavy mass state with $m_4 = 750$~keV and mixing $|U_{e4}|^2 = 2 \times 10^{-4}$ (just below the current upper limit at 95\% C.L.)~\cite{PhysRevLett.126.021803_beest} overlaid and shown in the inset. 

This detection principle, however, is generic in the sense that any isotope that can be contained in the nanosphere can be used, and the other isotopes in Table~\ref{tab:ec_isotopes} would produce similar sensitivity for $m_4$ up-to their respective $Q_{EC}$ values, as described in Sec.~\ref{sec:proj_sens_steriles}. However, among the isotopes in Table~\ref{tab:ec_isotopes}, $^7$Be provides a relatively unique case. Approximately 90\% of $^7$Be decays proceed to the $^7$Li ground state, with the emission of a single Auger $e^-$ with energy of $\sim 50$~eV~\footnote{The energy of the emitted electron in this light system depends on the chemical shift generated when it is sequestered in a material, as observed and noted in Ref.~\cite{PhysRevLett.125.032701}.}, and no accompanying X-rays. The single emitted Auger has a mean range in silica of $\lesssim5$~nm~\cite{ASHLEY1981127_sio2_stopping}, so in the vast majority of decays the Auger $e^-$ will stop in the nanosphere, providing a truly two body decay in which the total recoil momentum of the nanosphere is exactly equal and opposite to that of the outgoing $\nu$. While such a decay is especially simple to reconstruct, it has the drawback that there are no coincident secondary particles that can be used to veto backgrounds. For this reason, we consider below only the $\approx$10\% of $^7$Be decays that proceed through an excited state of $^7$Li, emitting a 478~keV $\gamma$ in coincidence. If backgrounds (see Sec.~\ref{sec:sterile_backgrounds}) could be made sufficiently low that this coincidence were not required for background rejection, $^7$Be decays to the ground state would present an especially simple system for sterile $\nu$ searches, possibly avoiding the need for the SiPM arrays surrounding the trap. 

\subsection{$\beta$ decays}
\label{sec:beta_decays}
\begin{figure}[t]
  \centering
  \includegraphics[width=\columnwidth]{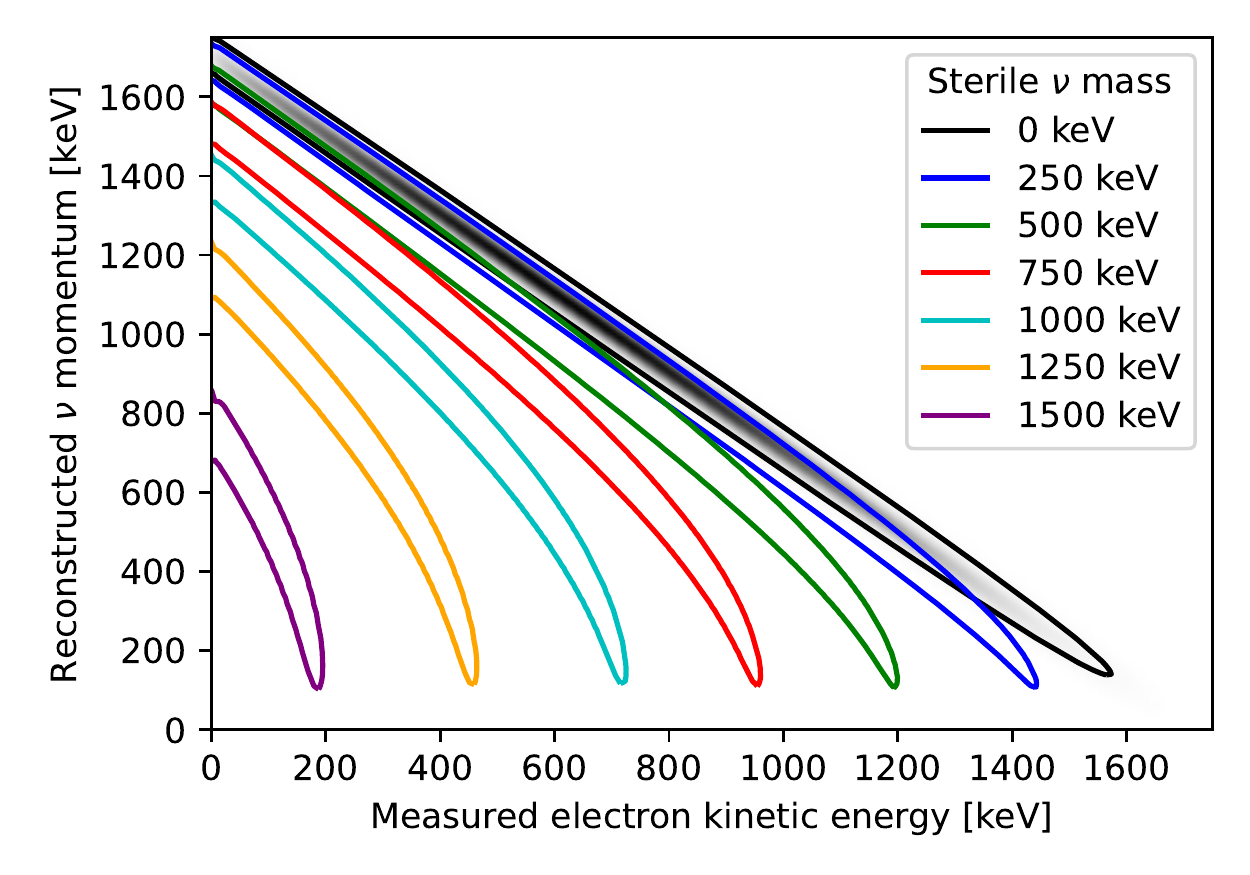}
  \caption{Calculated distribution of the reconstructed $\nu$ momentum, $|\mb{p}_\nu|$, versus electron kinetic energy, $T_e$, for $^{32}$P $\beta^-$ decay, for various values of $m_4$, assuming a 100~nm diameter nanosphere with the detection efficiencies described in the text. The probability density for this distribution for light neutrinos (with $m_i \approx 0$) is shown as the black shaded histogram, surrounded by the 2$\sigma$ contour containing 95\% of simulated events. The corresponding 2$\sigma$ contours for non-zero $m_4$ are indicated by the colored lines. }
  \label{fig:recoil_spec_beta}
\end{figure}

\begin{figure*}[t]
  \centering
  \includegraphics[width=\textwidth]{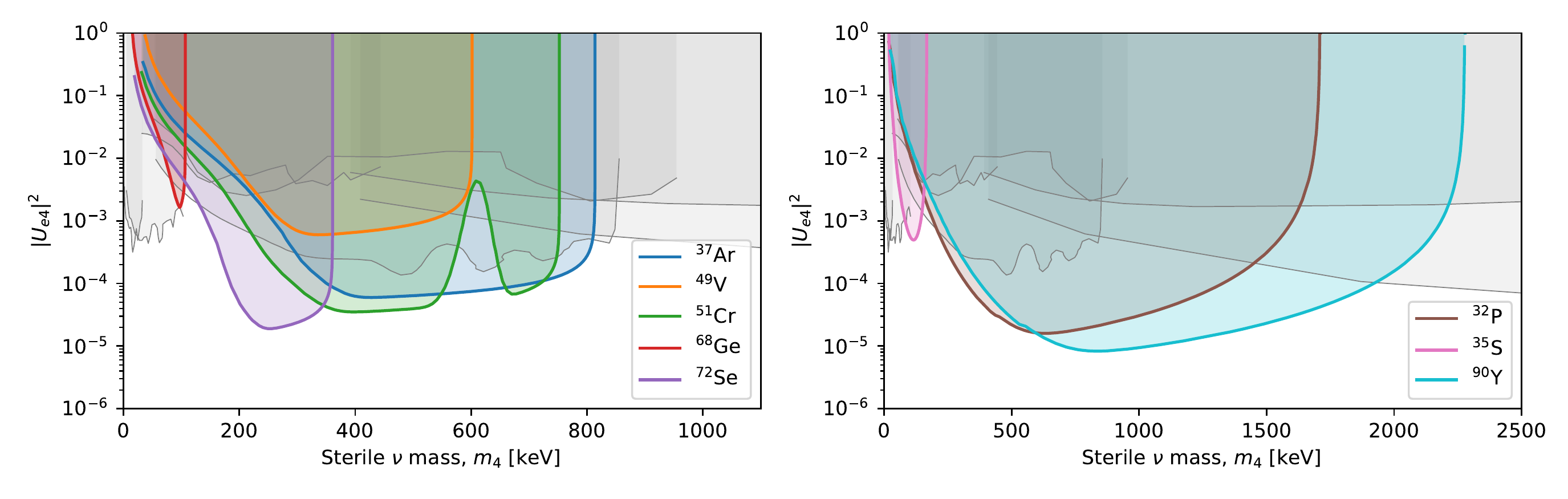}
  \caption{Projected sensitivity to the mixing of a heavy mass state, $|U_{e4}|^2$, as a function of mass, $m_4$, for a variety of isotopes, assuming a single, 100~nm diameter nanosphere is measured for 1~month with 1\% loading by mass of the isotope of interest, and assuming the detection efficiencies and experimental parameters described in the text. The sensitivity for isotopes decaying by EC (left) and $\beta^-$ decays (right) are shown separately. The projected sensitivity is compared to the most sensitive existing laboratory limits in this mass range (gray)~\cite{HOLZSCHUH1999247_ni63,HOLZSCHUH20001_s35,SCHRECKENBACH1983265_cu64, DEUTSCH1990149_f20, PhysRevLett.86.1978_re187, Kraus:2012he_mainz,Belesev:2013cba_troitsk,2018JETPL.108..499D_ce144pr144, PhysRevD.100.073011_pienu, PhysRevLett.126.021803_beest}. If sterile neutrinos constitute a significant fraction of dark matter, limits on X-ray emission from their radiative decays may also provide constraints in this mass range (not shown)~\cite{Dasgupta:2021ies_sterile_review}.}
  \label{fig:sensitivity_by_isotope}
\end{figure*}

For $\beta$ decays, in which both an $e^-$ and a $\bar{\nu}_e$ are present in the final state: 
\begin{equation}
\label{beta}
  \ce{^{A}_ZX} \rightarrow \ce{^{A}_{Z+1}Y} + e^- + \bar{\nu}_e,
\end{equation}
simply constructing the invariant mass of the neutrino through energy-momentum conservation (as done, e.g., in~\cite{2019NJPh...21e3022S_hunter_smith}), is not sufficient to optimally separate a population of decays emitting heavy sterile $\bar{\nu}$ from decays to light $\bar{\nu}$. Instead, optimal sensitivity can be reached through the correlation of the measured $\beta^-$ kinetic energy $T_e = E_e - m_e$ with the reconstructed neutrino 3-momentum. Energy-momentum conservation at the decay location gives the total neutrino energy $E_{\nu} = Q - T_e$, neglecting the tiny kinetic energy of the recoiling daughter nucleus. For an event with a given electron kinetic energy, the difference in the reconstructed neutrino momentum between a heavy $m_4$ state and effectively massless $m_i$ state is
\begin{align}
\label{betaendpoint}
|\mb{p}_4| - |\mb{p}_{i} | = \sqrt{(Q - T_e)^2 - m_4^2} - (Q-T_e)
\end{align}
The largest difference occurs for events where the electron gets the maximum allowed kinetic energy consistent with a final state $m_4$, $T_e \approx Q - m_{4}$, where $|\mb{p}_i| - |\mb{p}_{4}| \approx m_4$. Figure~\ref{fig:recoil_spec_beta} shows an example of the simulated reconstructed $|\mb{p}_\nu|$ versus $T_e$ for $^{32}$P, which decays to the ground state of $^{32}$S with $Q = 1710$~keV, for several different values of $m_4$. For low $T_e$, where the $\bar{\nu}$ carries energy much larger than $m_4$, the massive and approximately massless case converge for the lowest sterile $\nu$ masses considered. However, in the tail of the $\beta$ distribution at high $T_e$, the distributions can be distinguished as $m_4$ approaches $E_\nu$. 

\begin{table}[b]
    \centering
    \begin{tabular}{c | c c c}
Isotope &	$\beta$ endpoint [keV] & $T_{1/2}$ [day] & Live time [sphere days]\\
\hline
    $^3$H &	18.6 & 4500 & 252 \\
    $^{32}$P & 1711 & 14.3 & 2.8 \\
    $^{35}$S & 167 & 87.4 & 189 \\
    $^{90}$Y & 2279 (99.99\%) & 2.7 & 1.7 \\
    \end{tabular}
\caption{Selected $\beta^-$ decay isotopes considered here. The energy of the endpoint (i.e., maximum kinetic energy possible for the $\beta$ particle) and half-life are given for each isotope. In the case of $^{90}$Y, the decay to the ground state is dominant (99.99\% of decays). Rare decays to excited states of the $^{90}$Zr daughter are also possible, coincident with either $\sim$2~MeV conversion $e^-$ or de-excitation $\gamma$s. These decay branches are simulated in the MC, but overlap primarily with the low $T_e$ tail of the ground state decay and do not substantially affect the sensitivity at high $m_4$. The live time to observe $10^4$ decays is given following the assumptions in the text (i.e., 25~nm diameter polymer nanospheres for $^3$H, 50~nm diameter silica nanospheres for $^{35}$S, and 100~nm diameter silica nanospheres for $^{32}$P and $^{90}$Y). Isotope data are taken from~\cite{ENSDF}.}
    \label{tab:beta_isotopes}
\end{table}

Similar to EC, a number of candidate $\beta^-$-decay isotopes could be considered for such searches. Table~\ref{tab:beta_isotopes} provides some possible examples spanning a range of $Q$-values.  The complete list of possible $\beta^-$ emitters that may be suitable for such experiments is much larger than for EC given the number of bound neutron-rich systems in the lifetime range required, and the uniqueness of this weak decay mode.

\subsection{Projected sensitivity}
\label{sec:proj_sens_steriles}
The MC simulation described in Sec.~\ref{sec:steriles} is used to generate probability density functions (PDFs) for the reconstructed $\nu$ momentum, as a function of the neutrino mass, $m_4$. For EC decays, one-dimensional PDFs versus $|\mb{p}_\nu|$ are simulated, while two-dimensional PDFs of reconstructed $|\mb{p}_\nu|$ vs. $T_e$ are generated for $\beta$ decaying isotopes, similar to the examples shown in Figs.~\ref{fig:recoil_spec_EC} and \ref{fig:recoil_spec_beta}. For a given experimental live time, toy experiments are drawn from a light-$\nu$ only PDF ($m_i \approx 0$), given the $T_{1/2}$ for a given isotope and assuming a trigger efficiency for tagging the secondary particle of 40\% (limited by the solid angle coverage of the SiPM array). For experimental live times longer than $T_{1/2}$, we assume the nanospheres are replaced with new nanospheres containing the assumed initial isotope loading once every half-life (i.e., once every few days for the shortest half-lives considered), such that the number of decays remains roughly steady throughout the measurement period. A negative log likelihood (NLL) fit is then performed to each toy experiment, with a model consisting of a light $\nu$ plus heavy $\nu$ component (as a function of $|U_{e4}|^2$ at each $m_4$), and the median 95\% CL sensitivity for a given $m_4$ is determined~\cite{Workman:2022}. The resulting estimated sensitivities for a variety of EC and $\beta$ decaying isotopes with optimum sensitivity in the $m_4 \approx 0.1-1$~MeV range are shown in Fig.~\ref{fig:sensitivity_by_isotope}.

\begin{figure}[t]
  \centering
  \includegraphics[width=\columnwidth]{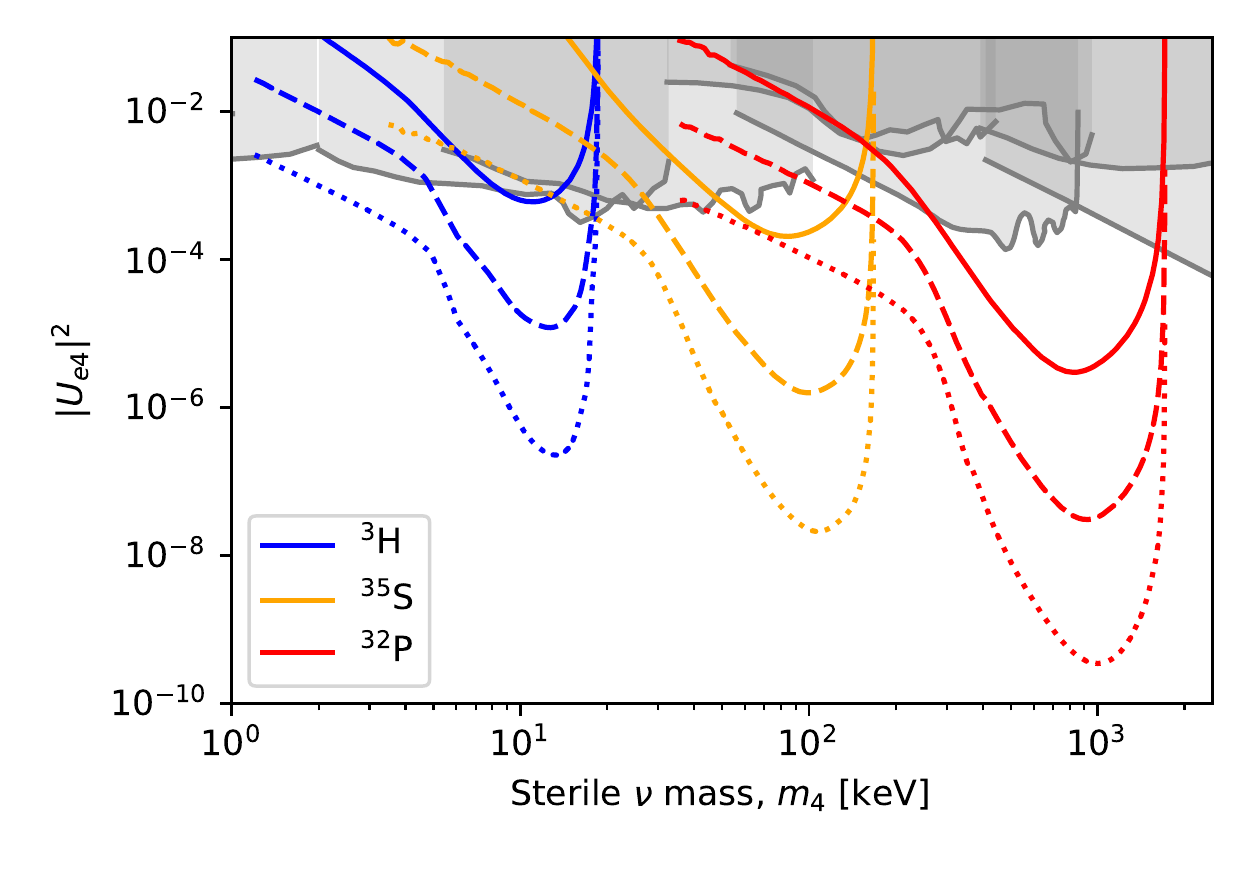}
  \caption{Estimated sensitivity for a variety of nanosphere sizes, exposures, and $\beta$-decay isotopes spanning a sensitivity range from $\sim$1--1000~keV. The background-free sensitivity for exposures of 1 nanosphere $\times$ 1 month (solid), 10 nanospheres $\times$ 1 year (dashed), and 1000 nanospheres $\times$ 1 year (dotted) are shown. The nanosphere sizes and loading fractions assumed for each isotope are described in the text. The most sensitive existing laboratory limits are also shown (gray)~\cite{HOLZSCHUH1999247_ni63,HOLZSCHUH20001_s35,SCHRECKENBACH1983265_cu64, DEUTSCH1990149_f20, PhysRevLett.86.1978_re187, Kraus:2012he_mainz,Belesev:2013cba_troitsk,2018JETPL.108..499D_ce144pr144, PhysRevD.100.073011_pienu, PhysRevLett.126.021803_beest}. If sterile neutrinos constitute a significant fraction of dark matter, limits on X-ray emission from their radiative decays may also provide constraints in this mass range (not shown)~\cite{Dasgupta:2021ies_sterile_review}.}
  \label{fig:sterile_sensitivity}
\end{figure}

As shown in Fig.~\ref{fig:sensitivity_by_isotope}, a large number of isotopes are expected to provide comparable sensitivity, covering the $\sim$100~keV to few MeV mass range. Measuring only a single 100~nm nanosphere for 1~month at existing detection limits would already probe up to an order of magnitude smaller values of $|U_{e4}|^2$ than existing searches, for the assumptions above. Several commonly used isotopes such as $^{32}$P provide significant coverage of the allowed parameter space. As shown in Fig.~\ref{fig:sterile_sensitivity}, larger exposures (either using arrays of nanospheres or longer integrations) can improve the sensitivity to smaller values of $|U_{e4}|^2$. 

Figures~\ref{fig:sensitivity_by_isotope}--\ref{fig:sterile_sensitivity} also show that existing sensitivities already demonstrated for 100~nm nanospheres are optimized for searching for $\sim$MeV scale mass sterile $\nu$. Although less developed to-date, sensitivity to lower masses may be possible with smaller nanospheres, which can reach correspondingly lower momentum thresholds. Similarly, reaching the SQL at lower trap frequencies can further improve the low mass reach. 

For example, Fig.~\ref{fig:sterile_sensitivity}, also shows the projected reach for searches with smaller nanospheres for lower endpoint isotopes such as $^3$H and $^{35}$S. For the lower momentum transfers for these isotopes, the nanosphere size and resonant frequency are tuned to maintain approximately a signal-to-noise ratio of 10 for recoils at the endpoint and SQL limited detection. For $^3$H, we assume a 25 nm diameter, $f_0=1$~kHz trap frequency, and a 20\% loading by mass that may be possible for a tritiated polymer (e.g. acrylic or polystyrene) nanosphere. For $^{35}$S, a 50~nm diameter silica nanosphere with $f_0 = 10$~kHz and 1\% mass loading is assumed. These lower resonant frequencies are typical of the expected frequencies in a radiofrequency (RF) trap at charge-to-mass ratios achievable for these smaller particles~\cite{Goldwater_2019_electro, PhysRevResearch.3.013018_northup_cooling}. While SQL limited detection for smaller particles allows lower momentum sensitivity, the reduced activity for a fixed number of nanospheres reduces the sensitivity to small branching ratios to heavy sterile $\nu$. These projected sensitivities are compared to $^{32}$P, using the 100~nm diameter silica nanosphere, $f_0 = 100$~kHz (corresponding to existing optical traps for this particle size), and 1\% loading also shown in Fig.~\ref{fig:sensitivity_by_isotope}. Higher trap frequencies or larger particles could be used while maintaining the same mass sensitivity if sub-SQL sensitivities were achieved (see further discussion in Sec.~\ref{sec:light_nu}).

\subsection{Backgrounds}
\label{sec:sterile_backgrounds}
The sensitivity projections in Figs.~\ref{fig:sensitivity_by_isotope}--\ref{fig:sterile_sensitivity} assume that the decays originating within the nanospheres can be identified with negligible backgrounds over the assumed exposure times. As with any rare event search, difficult to anticipate backgrounds are possible, and background studies will be required in prototype detectors to confirm the ultimate reach of such a system. However, reaching sufficiently low background levels for the projections above is plausible, as described in the following sections. 

\subsubsection{Recoil backgrounds}
Backgrounds can arise from interactions producing momentum recoils of the nanosphere at the keV--MeV scale, but which do not result from the decay of interest. However, the signature of a decay event can generally be tagged with a triple coincidence between the 1) nanosphere recoil; 2) detection of the secondary particle, e.g. the emitted $\beta$; and 3) detection of a change in the charge state of the nanosphere. We note that in the case of EC decays in which no Auger electrons escape the nanosphere and only X-rays are emitted, the third signature would be absent and only a double coincidence would be possible. However, for the EC decays considered in Table~\ref{tab:ec_isotopes}, with the exception of $^7$Be all decays would be expected to emit a $>$keV Auger $e^-$ in at least $\gtrsim 50$\% of decays. Events without a change in the charge of the nanosphere could be vetoed while maintaining most of the experimental live time.

Coincident detection of these three signatures would be expected to provide a background free signal even at the largest exposures assumed above, with the primary efficiency loss arising from the detection of the secondary particle (assumed to be 40\% efficient above). Further optimization of the experimental design might further improve this efficiency. Existing optical traps for microspheres in vacuum have measured spontaneous charging rates for neutralized particles below $1\ e^-$/week~\cite{Moore_2021,Monteiro2020uK}. Experiments searching for possible dark matter induced recoils of optically trapped particles found background recoil rates $<1$/day~\cite{Monteiro2020DM} at GeV momentum transfers, dominated by environmental acoustic and vibrational noise for which substantial further reduction is possible. With care in the choice of components and modest shielding around the detector, it should be possible to limit the rate of MeV-scale radiogenic and cosmogenic events detected by the SiPM array to $<10-100$~Hz (and much lower background rates have been achieved in underground detectors)~\cite{Schumann:2019eaa}. For a room temperature SiPM array with dark noise $\lesssim 1$MHz/mm$^2$, a detection threshold $>5$ photoelectrons (sufficiently low for tagging keV scale Auger $e^-$ or X-rays) would provide a sub-dominant background rate. Assuming these background rates are uncorrelated, the accidental coincidence rate for all three signatures within the 10~$\mu$s resolution of the event time would be as low as $10^{-11}$/(sphere yr), which is lower by many orders of magnitude compared to the exposures assumed above.

To generate a significant background, a nanosphere recoil rate of $\gg 1$~Hz at $\sim$MeV momentum transfers would be required. Due to the extremely small geometric cross-section of the nanospheres, external radiogenic and cosmogenic backgrounds provide orders-of-magnitude lower rates. Avoiding internal radiogenic decays would require minimizing contamination of the nanospheres by short-lived radioisotopes other than the species of interest. The maximum momentum transferred from gas collisions with the nanospheres is $<$100~keV for H$_2$ background gas at room temperature~\cite{Afek:2021vjy_coherent}, which would not provide a significant background for $m_4 \gtrsim 0.1$~MeV. For lower mass searches employing e.g. $^3$H or $^{35}$S, collisions with individual residual gas molecules could provide a significant background, but become negligible at pressures $\lesssim 10^{-11}$~mbar or through cooling of the chamber walls (and ambient gas temperature) to 4~K~\cite{Afek:2021vjy_coherent, Tebbenjohanns2021}. Other backgrounds considered for similar searches for nanosphere impulses in~\cite{Afek:2021vjy_coherent} are also expected to be sub-dominant for the searches proposed here.

\subsubsection{Containment of the recoiling nucleus}

If the nuclear recoil were to escape the nanosphere and fail to transfer its full momentum to the nanosphere COM, the reconstructed $\nu$ momentum can fall below the expected value when the recoil is fully stopped. Such events---if present for even a small fraction of decays---would lead to a background of events lying below the main distribution for decays emitting light neutrinos. While such events would not produce a well-defined peak at the expected momentum for decays to a heavy sterile state, they could obscure such a signal and limit the ultimate experimental sensitivity. 

Determining the fraction of such nuclear recoils at the relevant recoil energies that may escape the nanosphere is difficult. For $>$keV energies, the range of recoiling nuclei is well-characterized, and can be modeled in simulation tools such as SRIM-2013~\cite{2010NIMPB.268.1818Z_SRIM,srim_web}. SRIM calculations indicate that the mean projected range for the isotopes considered in Tables~\ref{tab:ec_isotopes}--\ref{tab:beta_isotopes} vary from 0.2--0.9~nm, with a standard deviation below 0.5~nm. However, we emphasize that SRIM has not been validated in detail at the relevant nuclear recoil energies from 3--70~eV, and these calculations require extrapolation to substantially lower energy than for which data exists. $^{32}$P, which is a light isotope with large endpoint energy, is expected to present the worst case for this background among the isotopes considered here. A SRIM simulation was performed for $^{32}$P in SiO$_2$ for nuclear recoil energies corresponding to electrons near the decay endpoint (i.e., $\sim$70~eV nuclear recoils), which gave a median projected range of 1.0~nm for the recoiling nuclei, with a maximum linear distance of 2.7~nm from the decay location for any of the $10^6$ simulated recoils. 

Using the SRIM calculations above as a rough estimate, approximately 5\% of recoils might be expected to escape a 100~nm diameter nanosphere with $^{32}$P uniformly distributed throughout its volume, which would present a potentially significant background if not otherwise vetoed (e.g., through identifying a change in the charge state of the nanosphere, or by detecting the emitted ion or atom). However, this calculation ignores the energy required to escape the nanosphere surface (due to chemical interactions, Van der Waals forces, etc), which may be relevant for the $\sim$eV scale recoil energies considered here. Dedicated measurements will likely be required to characterize this potential background for realistic nanospheres, over the range of recoil energies and charge states possible for various isotopes.

If such a background ultimately became a dominant limitation to the proposed techniques, a straightforward technical solution would be to deposit a few-nm thick shell of SiO$_2$ on the nanosphere exterior after fabrication of the nanosphere core containing the isotope of interest. Due to the standard ``bottom-up'' synthesis process for spherical nanoparticles of this size~\cite{1968JCIS...26...62S_stober}, such a layer can be chemically accreted after production of the core, and core-shell silica nanospheres of exactly this type containing a $\beta$ emitting core surrounded by an inactive shell are commercially available~\footnote{Corpuscular Inc., \url{http://www.microspheres-nanospheres.com/}}. Similar particles are also commonly produced for medical applications when toxicity from the escape of the recoiling nuclei is a concern~\cite{llop2016isotopes_in_nano}.

\subsubsection{Misreconstruction of secondary particles}
For searches with $\beta$ decaying isotopes, misreconstruction of the kinetic energy of the $e^-$ can lead to errors in the estimate of the $\bar{\nu}$ momentum and inferred mass. For example, $e^-$ that backscatter from the scintillator and do not deposit their full energy could be reconstructed at lower $T_e$, falling below the main distribution for well-reconstructed events. Such events could present a similar background distribution to that described above for events where the nuclear recoil may escape the nanosphere. Optimizing the detector design to avoid (or detect and veto) such backscattered events is important for reaching the background free sensitivities above. Other sources of event misreconstruction due to e.g. detector gaps or edge effects, noise, etc, must also be mitigated. However, if sufficiently accurate reconstruction of secondary $e^-$ ultimately proves infeasible at the desired background levels, selecting EC decays emitting only X-rays (with a measurement of the charge state ensuring no Auger $e^-$ escaped the nanosphere) would avoid any need for accurate reconstruction of the secondary energy for sterile neutrinos in the $\gtrsim 100$~keV mass range.

\section{Towards light neutrino mass measurements}
\label{sec:light_nu}
While the searches for heavy mass states described in the previous section appear to be feasible with existing optomechanical technologies, an obvious question is whether such levitated mechanical systems could ultimately be extended to reach sensitivity to the light masses, $m_{i}$ ($i = 1,2,3$), of which at least two are non-zero. The best existing laboratory constraints indicate that the mass scale of these light $\nu$ is $< 800~{\rm meV}$~\cite{KATRIN:2021uub_sub_ev}, while cosmological measurements based on fits to the standard cosmological model suggests the sum of masses is $\lesssim 200~{\rm meV}$~\cite{Planck:2013pxb_nu_mass, DES:2021wwk_nu_mass}. Neutrino oscillation experiments provide a lower bound on the mass scale, requiring at least one of the light $\nu$ to have mass $\gtrsim 50$~meV~\cite{Esteban:2020cvm_nufit}. Given the much smaller masses than the heavy sterile case, reaching sensitivity to the masses of the SM $\nu$ is substantially more difficult. Indeed, directly measuring the masses of the SM $\nu$ is the focus of a decades long experimental effort involving large scale detectors~\cite{Formaggio:2021nfz}, and is key to identifying the beyond the SM processes that may account for the origin of the non-zero masses of the light states.

An ambitious goal for these techniques would be to perform momentum reconstruction sensitive enough to resolve the masses of the light neutrinos on an event-by-event basis, which we focus on below. However, a nearer-term measurement might be to statistically detect the scale of neutrino masses, averaging over a large number of decays $N$. With a $\beta$ decay, even in the most optimistic scenario and working near the endpoint, the first goal requires momentum measurements with sensitivity around the light neutrino masses $\Delta p \approx m_i$ (see Eq. \eqref{betaendpoint}), i.e. at least as good as $\Delta p \lesssim 200~{\rm meV}$. If systematic errors were reduced sufficiently to permit averaging over many events, measurement of the effective mass $m_\beta = \sqrt{\sum_{i=1}^3 |U_{e i}|^2 m_i^2}$~\cite{Formaggio:2021nfz} could be achieved with $\approx \sqrt{N}$ poorer event-by-event sensitivity. We discuss what would be required for an optomechanical sensor to achieve these sensitivities below.

Beyond the required momentum sensitivity, there are several conceptual challenges that must be overcome for possible measurements of the light neutrino masses:
\begin{enumerate}[label=(\Alph*)]
    \item \label{Qvals} For typical EC or $\beta$ decay energies, the light $\nu$ are highly relativistic and the effect of their rest mass on their momentum is correspondingly suppressed. One needs to find a decay with much lower energy release $Q$ and work near the end of the spectrum where the $\nu$ are emitted with low kinetic energy.
    \item \label{heisenberg} Heisenberg uncertainty requires that the initial position uncertainty $\Delta x$ of the decaying parent is sufficiently large to be consistent with the small required momentum uncertainties.
    \item \label{broadening} Secondary processes due to the coupling of the decaying atom to its surroundings can significantly broaden the momentum carried by an emitted $\nu$ in solid materials.
\end{enumerate}
The latter two issues have received considerable attention recently, because they may cause fundamental difficulties for experiments aiming to measure neutrino masses and relic neutrino capture events \cite{Cheipesh:2021fmg,PhysRevD.105.043502,PTOLEMY:2022ldz_theory}. The localization of the parent nucleus (B) has also been recently considered in the context of neutrino oscillations of reactor neutrinos~\cite{PhysRevD.106.053007_osc_loc, deGouvea:2021uvg_osc_loc, Akhmedov:2022bjs_osc_loc, Jones:2022cvh_osc_loc}. While the broadening issue \ref{broadening} will require detailed solid state engineering, the momentum reconstruction technique described here, combined with quantum control of the center-of-mass of the solid in which the isotopes are located, may provide new tools for solving issue \ref{heisenberg}.

\subsection{Ultra-low $Q$ value decays}
As described above, kinematic reconstruction of the neutrino mass is most effective for non-relativistic neutrinos (see Eqs. \eqref{eq:momentum_diff} and \eqref{betaendpoint}). For typical EC $Q$-values and $\beta$ decay endpoints with $E_\nu \approx$~MeV, the corresponding difference in momentum sensitivity needed to resolve the neutrino mass in a given event $\Delta|\mb{p}_\nu| \lesssim 1~ {\rm \mu eV}$, even at the current laboratory upper limits on $m_i$. The momentum difference could be made significantly larger, up to $\Delta |\mb{p}_{\nu}| \approx m_i \lesssim 200~{\rm meV}$, by observing much lower energy $\nu$ than in typical nuclear decays. 

Common ``low-energy'' decays such as $^3$H ($Q = 18.6$~keV) and $^{163}$Ho ($Q = 2.8$~keV) would still require very large arrays of nanoparticles, since decays emitting $\nu$ with energies $E_\nu \lesssim 1$~eV make up only a tiny fraction of the overall decay rate~\cite{Formaggio:2021nfz}. There is, however, significant interest in identifying a new class of ``ultra-low'' $Q$ value EC~\cite{PhysRevC.106.015501, PhysRevLett.127.272301_dy159} or $\beta$~\cite{Gamage:2019xvx_ultra_low} decays, in which a branch of the decay occurs to an excited state of the daughter. If the energy of this excited state roughly corresponds to the total transition energy, an accidental cancellation is possible, providing $Q<1$~keV. A known example of such a transition is in $^{115}$In~\cite{PhysRevLett.103.122502_in115, PhysRevLett.103.122501_in115}, which typically decays via $\beta^-$ emission with an end-point energy of $\approx 0.5$~MeV, but for which a small branching ratio ($\sim 10^{-6}$) to an excited state exists, providing a $\beta^-$ transition with an endpoint of $147$~eV. However, the long half-life of this decay ($4.4 \times 10^{14}$~yr) and small branching ratio to the excited state transition translates to a much larger required mass per decay with $E_\nu \lesssim 1$~eV than $^3$H or $^{163}$Ho~\cite{Formaggio:2021nfz}. 

Nonetheless, a recent survey has identified more than 100~candidates for other potential ultra-low $Q$ value $\beta$ or EC transitions via excited states, which remain to be precisely measured~\cite{Gamage:2019xvx_ultra_low}. Realizing an optomechanical measurement of the light $\nu$ masses would likely require a transition with sufficiently low $Q$ value, experimentally manageable $T_{1/2}$, and high branching ratio to be identified through such measurements. While speculative, the existence of such a transition is plausible, and we strongly encourage the precision atomic mass measurement community to continue their work in this area.  We further comment on the measurement requirements below if such a transition were identified. Of course, such a transition may permit other existing techniques to be used to study neutrino masses~\cite{Formaggio:2021nfz}, although here we focus only on the unique properties of an optomechanical measurement.

\subsection{Quantum state preparation and readout sensitivity}
The measurement sensitivity needed to resolve the light neutrino masses on an event-by-event basis $\Delta \mb{p} \lesssim 200~{\rm meV}$ entails significant constraints on both the initial state of the source and measurement process from quantum mechanics. Comparing to \eqref{sql}, we see that obtaining sufficient momentum sensitivity will require some combination of using smaller mass mechanical systems, low-frequency traps, and/or techniques like squeezing \cite{aasi2013enhanced,McCuller2020,Gonzalez-Ballestero:2022ktw_squeeze} or backaction evasion \cite{Braginsky:1990ei,Danilishin:2017ndl,Ghosh:2019rsc} to get below the SQL. 

For concrete examples, we can consider the requirements to obtain the $\Delta p$ sensitivities above. With a silica nanosphere of radius $10~{\rm nm}$ trapped at $2\pi \times 1~{\rm kHz}$, the SQL corresponds to $\Delta p \approx 50~{\rm eV}$, so measurements at the SQL should---at least in principle---be capable of resolving a non-zero $m_\beta$ by averaging over $N \approx 10^6$ decays if limited only by statistical noise. To reach $\Delta p \sim 100~{\rm meV}$ sensitivity, one could use a monolayer graphene flake with area $25~{\rm nm}^2$, levitated in a weakly confining trap with frequency $\omega = 2\pi \times 100~{\rm Hz}$, although a significant amount of squeezing to achieve sensitivity $20~{\rm dB}$ below the SQL is required even with these parameters. While challenging, performing such sub-SQL measurements is a key objective of the optomechanics community in the coming decades \cite{Levitodynamics2021}, and other recently proposed experiments require higher levels of squeezing and detection of smaller momentum transfers than required here~\cite{Bose2017, Marletto2017,Aspelmeyer:2022fgc}.

\begin{figure}[t]
  \centering
  \includegraphics[width=\columnwidth]{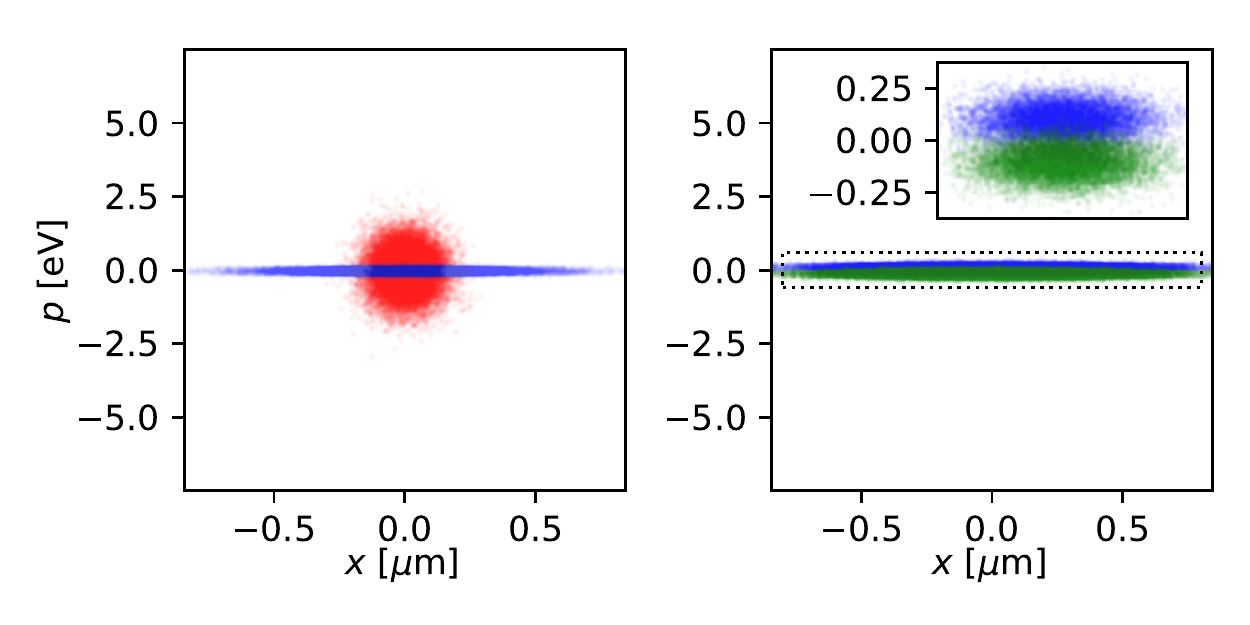}
  \caption{ Qualitative example of requirements for the nanoparticle state preparation and measurement following the decay. (left) Comparison between the quasiprobability distribution for the zero point motion of the COM of a 5~nm diameter graphene flake cooled to the ground state of a weakly confining harmonic potential ($\omega = 2\pi \times 100$~Hz) [red], and after squeezing the momentum quadrature of the nanoparticle's motional state by 20~dB [blue]. (right) Quasiprobability distribution following the decay for the nanoparticle COM (blue) and neutrino (green) assuming all secondary particles are trapped within the flake, and the $\nu$ carries momentum $|\mb{p}_\nu|$=100~meV. (inset) Zoom to region highlighted by black dotted box in the main figure.}
  \label{fig:squeeze}
\end{figure}

The optomechanical strategy proposed here also has fundamental advantages over other techniques. In particular, if the unstable isotope is bound to a fixed substrate, it should be localized in position space to within atomic distances $\Delta x \lesssim 1~{\rm nm}$ \cite{Cheipesh:2021fmg}. Heisenberg uncertainty then implies an initial-state momentum uncertainty $\Delta p \gtrsim (2~{\rm nm})^{-1} \approx 100~{\rm eV}$. This would naively render the identification of the neutrino masses impossible. However, the ability to monitor the center-of-mass DOF of the mechanical element with sufficient accuracy $\Delta p$ to detect the neutrino masses means in particular that it can be prepared in a sufficiently narrow momentum-space wavepacket~\footnote{Note that the mechanical system is only very weakly trapped $\omega_s \ll 1~{\rm MHz} \approx 60~{\rm neV}$, and so during the decay event the entire system is essentially freely-falling.}. This corresponds to delocalizing the entire system in position space. Such delocalization requires, e.g., cooling the nanoparticle to the ground state of the trapping potential and squeezing the motion of the nanoparticle to delocalize its position and narrow its momentum distribution (see Fig.~\ref{fig:squeeze}). While optimal protocols for this application are the subject of future work, levitated optomechanical systems provide substantial ability to control the optical potential in real time, allowing the preparation of squeezed or highly non-Gaussian states (see, e.g.,~\cite{PhysRevLett.127.023601_deloc,Neumeier:2022czd_squeeze} for recent proposals of experimental protocols similar to those that would be required here).

We also emphasize that due to momentum conservation, the internal states of the nanoparticle do not need to be directly controlled or measured. It is sufficient to cool, prepare, and measure only the COM motion of the nanoparticle. For example, the decaying nucleus will initially be in a thermal motional state near room temperature, but this internal DOF does not affect the nanoparticle COM and the $\nu$ leaving thenanoparticle will remain entangled with the COM motion such that a sufficiently precise measurement of the nanoparticle momentum can determine the outgoing $\nu$ momentum~\footnote{We note that there are in principle small corrections to the momentum measured in the nanoparticle COM frame relative to the rest frame of the decaying nucleus that depend on the thermal velocity of the nucleus, internal excitations, etc., but which are suppressed relative to the momentum of the emitted particle by the mass ratio of the emitted particle to the nuclear mass. However, due to the maximum practical energy available in nuclear decays (10--20~MeV), and the minimum nuclear masses $>$GeV, these would typically be expected to provide sub-dominant corrections to the inferred momentum in practical experiments}. In Appendix \ref{manybody}, we illustrate the above points with a simple toy model consisting of the parent nucleus harmonically bound to a solid object.

\subsection{Spectral broadening effects}
A key advantage of the techniques proposed here is the ability to concentrate the isotope of interest in a solid matrix, permitting high densities of the isotope in a small volume. However, interactions between the atoms of the decaying isotope and the host material can lead to broadening of the transitions of interest at the $\sim$eV scale due to the effect on the electronic binding energies in the parent and daughter atoms on the total energy available to the decay (analogous to ``inhomogeneous broadening'' in solid state laser systems). While not a fundamental limitation, a solid-state source for such decays would likely require the isotope to be positioned in a highly controlled, crystalline lattice to avoid such inhomogeneous broadening~\cite{Samanta_arxiv_materialeffects}. Similar ideas have been proposed in the literature to create sources based on loading graphene~\cite{PTOLEMY:2019hkd, PTOLEMY:2018jst,Betti:2018bjv,Apponi:2021hdu} or other host crystals~\cite{Raghavan:2005gn_moss, Raghavan:2006xf_moss} with $\beta$ emitters. We note that levitation of $\mu$m sized graphene flakes has been demonstrated in RF traps~\cite{PhysRevB.82.115441_kane_graphene}, and these traps could be used to confine other nanoscale objects of interest~\cite{Goldwater_2019_electro, PhysRevResearch.3.013018_northup_cooling}.

\subsection{Experimental feasibility}
To perform such a measurement, we consider the possibility that an experimentally viable $\beta$ transition to an excited state of the daughter with an endpoint of $\sim$100~eV is identified from the list of candidates in~\cite{Gamage:2019xvx_ultra_low}. Such a transition will provide a spectrum of decays extending to $E_\nu \lesssim 1$~eV, which will provide the dominant sensitivity to $m_i$. Ideally this isotope decays through an allowed transition to a closely spaced excited state, with a partial decay constant to the relevant excited state branch $\lambda \gtrsim 1~$~yr$^{-1}$ to enable a sufficient number of decays to be observed, even for nanoparticles of the mass assumed in Fig.~\ref{fig:squeeze}. Finally an overall half-life long enough that the isotope can be produced and implanted into the nanoparticles is needed. While not in principle required, it is also preferable that the excited state decay to the ground state of the daughter proceeds through a forbidden transition, leading to a delay in the emission of the high-energy de-excitation $\gamma$s to allow separation of the much smaller momentum carried by the $\nu$. These $\gamma$s can still be triggered on to avoid backgrounds, while temporally distinguishing the momentum transferred to the nanoparticle. 

\begin{figure}[t]
  \centering
  \includegraphics[width=\columnwidth]{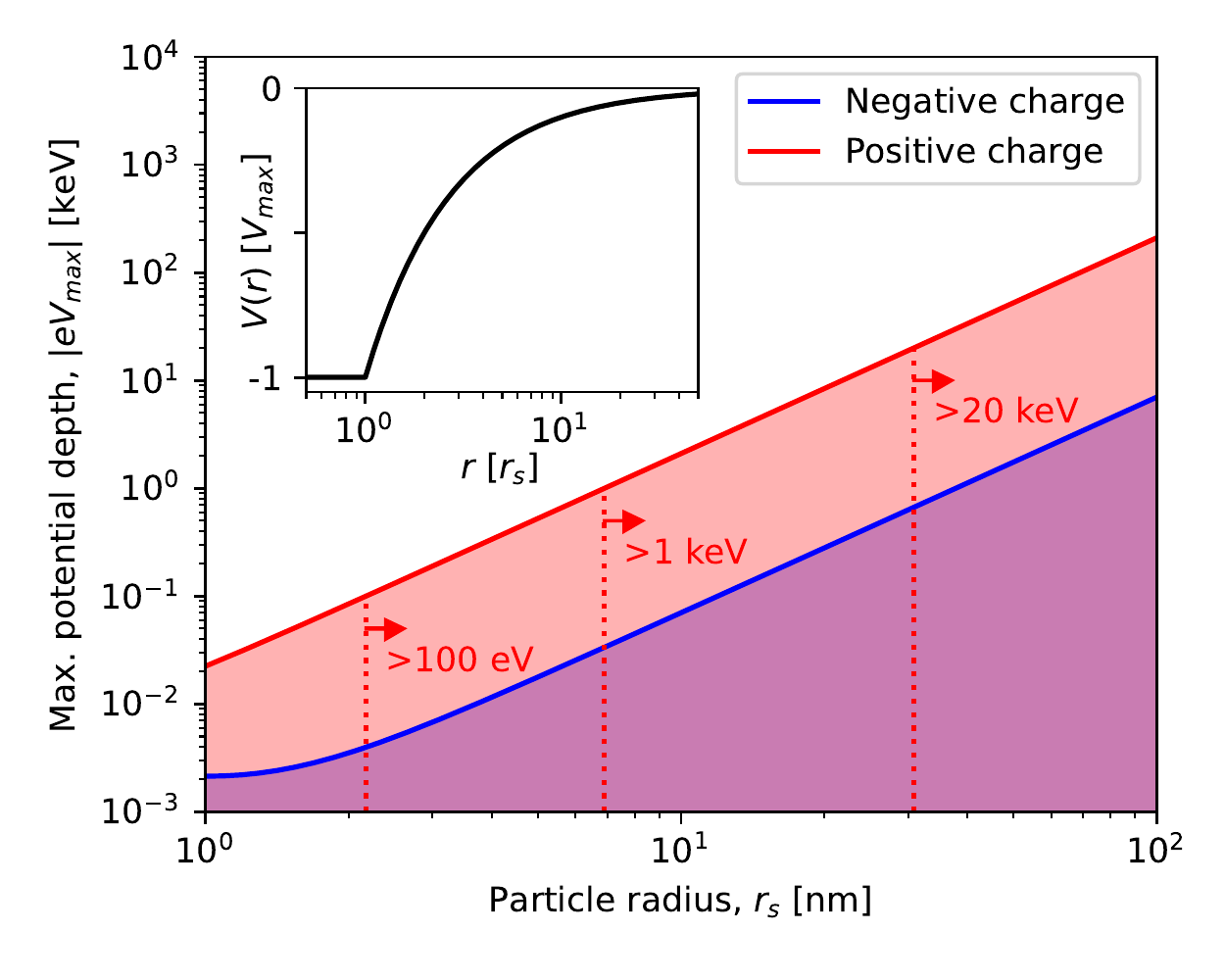}
  \caption{ Maximum depth of the electric potential energy, $V_{max}$, for a positively (red) or negatively (blue) charged nanosphere of radius $r_s$, assuming the net charge is limited by field emission~\cite{Goldwater_2019_electro, 1987ApJ...320..803D_charging}. For $\beta^-$ decay, the red dashed lines and arrows indicate the nanosphere sizes for which an electron of a given initial kinetic energy could be stopped by a nanosphere charged to this limit. (inset) Electric potential as a fraction of $V_{max}$ relative to a distant grounded surface as a function of distance from the nanosphere center, $r$. }
  \label{fig:charge_vs_rad}
\end{figure}

For such a $\beta$ decay, precise measurement of the $e^-$ momentum at the required precision is technologically challenging, although the point-particle nature of the decay source may reduce the challenges related to focusing and collecting the $e^-$ for, e.g., a time-of-flight measurement~\cite{Formaggio:2021nfz, Carney2021Trapped}. An alternative possibility is to positively charge the nanoparticle to provide precise control of the energy of any $e^-$ leaving the nanoparticle and limit its momentum. In particular, charging the nanoparticle itself to a $\gg$100~V potential relative to the surrounding electrodes can ensure that the $e^-$ is recaptured by the particle itself, and the final state is a true 2-body decay with an outgoing $\nu$ and recoiling nanoparticle only. Field emission (i.e., ejection of electrons or ions from a solid nanoparticle due to the electrostatic field at its surface) limits the maximum positive charge on an nanoparticle of radius, $r_s$ to $Q/e \approx 21(r_s/[1\ \mathrm{nm}])^2$~\cite{Goldwater_2019_electro, 1987ApJ...320..803D_charging}, permitting sufficient potentials to prevent the emission of the $e^-$ to be reached on a $\gtrsim 5$~nm diameter nanoparticle (see Fig.~\ref{fig:charge_vs_rad}). The charge on the nanoparticle could also be tuned with single $e^-$ precision~\cite{Moore2014, Frimmer2017}, to allow only the highest energy $e^-$ to escape and be tagged (and thus identify only the lowest energy $\nu$). While calculating in detail the spectrum of emitted $e^-$ from such a graphene flake as a function of its net charge is beyond the scope considered here, measurements of the recoil spectrum versus charge may provide a precise experimental method for identifying the effect of a non-zero $\nu$ mass. 

\section{Outlook}
We have described the use of mechanical sensors whose motional degrees of freedom are controlled and measured in the quantum regime to kinematically reconstruct the mass of neutrinos emitted in weak nuclear decays. Such systems offer a new set of tools for searching for a variety of invisible particles that may be emitted from nuclei---including massive neutrinos---and offer several advantages over existing techniques. In particular, trapped nanoscale objects permit a variety of isotopes to be characterized without, e.g., the need to demonstrate laser cooling of a given species, while reaching sensitivities that are sufficient to resolve the momentum in a single nuclear decay. Solid materials allow a high density of nuclei to be confined in a trap, and enable control and readout of the motional state of the particle using tools from quantum optomechanics. 

The estimates performed here indicate that even a single trapped nanosphere operating at existing sensitivity levels could probe new parameter space for $\sim$MeV scale sterile neutrinos in a few days integration, if $\beta$ or EC decaying nuclides were embedded in the particle at percent level concentrations. Future extensions to large arrays of such particles may be able to reach many orders-of-magnitude lower branching ratios for such decays, for a variety of isotopes spanning the keV to MeV mass range. While well beyond the current state of the art, it is plausible that smaller particles may eventually reach the momentum sensitivity needed to detect the mass of the light SM neutrinos. If an ultra-low $Q$ value EC or $\beta$ transition were also identified with sufficiently high decay rate, detection of the light neutrino masses with similar ideas proposed here may one day become possible.

\begin{acknowledgments}
We thank Maurice Garcia-Sciveres, W. Linda Xu, and Giorgio Gratta for discussions. This work was initiated at the Aspen Center for Physics, which is supported by NSF Grant PHY-1607611. DC is supported by the US Department of Energy under contract DE-AC02-05CH11231 and Quantum Information Science Enabled Discovery (QuantISED) for High Energy Physics grant KA2401032. DCM is supported, in part, by the Heising-Simons Foundation, NSF Grants PHY-1653232 and PHY-2109329, and ONR Grant N00014-18-1-2409.  KGL is supported, in part, by the Gordon and Betty Moore Foundation and the US DOE-SC Office of Nuclear Physics under grants DE-FG02-93ER40789 and DE-SC0021245.
\end{acknowledgments}

\bibliography{sterile_nus.bib}

\appendix

\section{Toy many-body model of the decay}
\label{manybody}

Consider a three-body system consisting of a heavy mass $H$, a light mass $L$ bound to $H$, and a neutrino $\nu$. Here the heavy mass $H$ represents our solid sphere, the light mass $L$ represents the unstable parent isotope, and we will ignore any secondary decay products beside the nucleus. The Hamiltonian is
\be
H = H_H + H_L + H_{\nu} + \frac{1}{2} k (\mb{x}_H - \mb{x}_L)^2 + V_{L\nu}.
\ee
The kinetic energy of the heavy system $H_H = \mb{p}_H^2/2 m_H$. We have modeled the binding of the isotope $L$ to the rest of the sphere $H$ as a harmonic potential, but in what follows the exact form of the potential is unimportant. We model the isotope $L$ as carrying its momentum as well as a internal two-state system $\ket{\mb{p}_L,a}$ with $a \in \{ e,g \}$ representing the nuclear excited and ground states relevant to our decay, so
\be
H_L = \sqrt{\mb{p}_L^2 + (m_L + \delta_{ae} Q)^2}.
\ee
Finally we model the neutrino as carrying momentum as well as an internal three-state system $\ket{\mb{p}_{\nu},b}$ with $b \in \{ 0, \ell, h \}$ representing either no neutrino, a light neutrino mass eigenstate $\ell$, or a heavy neutrino eigenstate $h$, so
\be
H_{\nu} = \begin{pmatrix} 0 & & \\ & \sqrt{\mb{p}_{\nu}^2 + m_{\ell}^2} & \\ & & \sqrt{\mb{p}_{\nu}^2 + m_{h}^2} \end{pmatrix}.
\ee
Finally, the interaction potential $V_{L\nu}$ depends only on the relative position $V_{L\nu} = V_{L\nu}(\mb{x}_L - \mb{x}_{\nu})$, with matrix elements of the form
\be
V_{L\nu} = V_{\ell} \ket{g \ell} \bra{e 0} + V_{h} \ket{g h} \bra{e 0},
\ee
which models the decay of the excited nuclear state to its ground state plus an ejected neutrino. Here and throughout this paper we follow the conventions of \cite{Weinberg:1995mt}, so in particular the metric has $-+++$ signature and momentum eigenstates are normalized as $\braket{\mb{p} | \mb{p}'} = \delta^3(\mb{p}-\mb{p}')$.

This Hamiltonian conserves the total momentum
\be
[H,\mb{p}_{\rm total}] = 0, \ \ \ \mb{p}_{\rm total} = \mb{p}_H + \mb{p}_L + \mb{p}_{\nu}
\ee
since it is only a function of pairs of relative coordinates and $[\mb{p}_{\rm total},\mb{x}_i - \mb{x}_j]=0$. We can also define the total (center of mass) and relative momenta of the $HL$ bound system,
\begin{align}
\label{comcoords}
\mb{P} = \mb{p}_H + \mb{p}_L, \ \ \ \mb{p} = \mu \left( \frac{\mb{p}_H}{m_H} - \frac{\mb{p}_L}{m_L} \right)
\end{align}
where as usual the reduced mass is $\mu = m_H m_L/(m_L + m_H)$. The total momentum $\mb{P}$ in this bound system is not a conserved quantity, but we can rewrite total momentum conservation as
\begin{align}
\label{momentumcons}
[H,\mb{P}] = - [H,\mb{p}_{\nu}].
\end{align}
In other words, if the nucleus decays and a neutrino is ejected with momentum $\mb{p}_{\nu}$, then correspondingly the momentum of the bound system changes $\Delta \mb{P} = - \mb{p}_{\nu}$. Notice that the relative momentum $\mb{p}$ does not enter the total spatial momentum $\mb{p}_{\rm total}$; it is a purely internal degree of freedom. Our goal is to illustrate how the total three-body system responds to the decay of the nucleus $L$. In particular, we assume that in practice we only have experimental access to the center-of-mass coordinate $\mb{X} = (m_H \mb{x}_H + m_L \mb{x}_L)/(m_H + m_L)$, but not separately $\mb{x}_H$ or $\mb{x}_L$. 

To begin, consider an initial state of the total system where both the COM and relative bound state momenta are in exact eigenstates, the nucleus is excited, and there is no neutrino:
\be
\ket{\psi} = \ket{\mb{P},\mb{p},e; 0}.
\ee
We can equivalently write this as an eigenstate in the $\ket{\mb{p}_H, \mb{p}_L}$ basis. This latter representation is convenient to represent the decay event. We make a key assumption, which is that the decay happens instantaneously compared to the timescales internal to the $HL$ system. This means that an initial $L,\nu$ state $\ket{\mb{p}_L,e; 0}$ evolves to
\begin{align}
\begin{split}
& \int d^3 \mb{p}_{\nu}' \Big[ \mathcal{M}_{\ell} \delta(E_L - E_L' - E^{\ell}_{\nu}) \ket{\mb{p}_L - \mb{p}_{\nu}',g; \mb{p}_{\nu}', \ell} \\
& + \mathcal{M}_{h} \delta(E_L - E_L' - E^h_{\nu}) \ket{\mb{p}_L-\mb{p}_{\nu}',g; \mb{p}_{\nu}', h} \Big],
\end{split}
\end{align}
while leaving the state of the $H$ system unchanged. Here the outgoing neutrino has momentum $\mb{p}_{\nu}' = \mb{p}_L - \mb{p}_L'$ by momentum conservation. The arguments of the energy-conservation delta functions can be approximated as $Q - \sqrt{\mb{p}_{\nu}'^2 + m_{\ell,h}^2}$, respectively, under the assumption that $\mb{p}_L, m_{\ell,h}, Q \ll m_L$, which is certainly true in our problem. Here $\mathcal{M}_{\ell,h}$ represent probability amplitudes for the decay into the various neutrino mass states; in our problem these are proportional to $U_{e\ell} G_F$ and $U_{eh} G_F$ respectively, where $G_F$ is the Fermi constant. In general, $\mathcal{M}_{\ell,h}$ can depend on $\mb{p}_{\nu}'$ as well as the initial conditions; in the relevant regime for nuclear decays this dependence is very weak and we will just take $\mathcal{M}_{\ell,h}$ to be constants. Moving back to total and relative coordinates, we have the final state
\begin{align}
\begin{split}
\label{finalpure}
& \int d^3 \mb{p}_{\nu}' \Big[ \mathcal{M}_{\ell} \delta(Q - \sqrt{\mb{p}_{\nu}'^2 + m_{\ell}^2}) \ket{\mb{P} - \mb{p}_{\nu}', \mb{p}+\mb{p}_{\nu}'; \mb{p}_{\nu}', \ell} \\
& + \mathcal{M}_{h} \delta(Q - \sqrt{\mb{p}_{\nu}'^2 + m_{h}^2}) \ket{\mb{P} - \mb{p}_{\nu}', \mb{p}+\mb{p}_{\nu}'; \mb{p}_{\nu}', h} \Big].
\end{split}
\end{align}
To get this result we used the approximation $m_L \ll m_H$, so that $\mb{p} \approx -\mb{p}_L$, and everywhere the internal nuclear state is understood to be $\ket{g}$.

Finally, let us consider a more realistic initial state. We assume that the relative coordinate ``internal mode'' $\mb{p}$ is prepared in a general density matrix,
\be
\rho_{\rm rel} = \int d^3\mb{p} d^3 \ul{\mb{p}} \rho_{\rm rel}(\mb{p},\ul{\mb{p}}) \ket{\mb{p}} \bra{\ul{\mb{p}}}.
\ee
In practice, this is something like a thermal state at room temperature. Unlike the relative mode, we have experimental access to the COM motion $\mb{P}$, and therefore can assume we prepare it in a particular pure state 
\be
\ket{\psi_{\rm COM}} = \int d^3\mb{P} \psi(\mb{P}) \ket{\mb{P}}.
\ee
For example, with SQL-level readout and control, this could be the ground state in an optical harmonic potential. For sub-SQL readout and control, as needed for example in a measurement of the light neutrino masses, this state could be taken as a squeezed state with momentum variance below the scale of the ground state of the trap $\Delta P \ll \Delta P_{SQL}$ (see Eq. \eqref{sql}). 

Our total initial state in this case is now
\be
\rho = \ket{\psi_{\rm COM}} \bra{\psi_{\rm COM}} \otimes \rho_{\rm rel} \otimes \ket{0} \bra{0}.
\ee
The post-decay density matrix consists of superpositions of the form \eqref{finalpure}, weighted by the appropriate initial wavefunction and density matrix elements:
\begin{align}
\begin{split}
& \rho_f = \int d^3\mb{P} d^3\ul{\mb{P}} d^3\mb{p} d^3\ul{\mb{p}} d^3\mb{p}_{\nu} d^3\ul{\mb{p}}_\nu \psi(\mb{P}) \psi^*(\ul{\mb{P}}) \rho_{\rm rel}(\mb{p},\ul{\mb{p}})  \\
& \times \Big[ \mathcal{M}_{\ell} \delta(Q - \sqrt{\mb{p}_{\nu}^2+m_{\ell}^2}) \ket{\mb{P} - \mb{p}_{\nu}, \mb{p} + \mb{p}_{\nu}; \mb{p}_{\nu}, \ell} \\
& + \mathcal{M}_{h} \delta(Q - \sqrt{\mb{p}_{\nu}^2+m_{h}^2}) \ket{\mb{P} - \mb{p}_{\nu}, \mb{p} + \mb{p}_{\nu}; \mb{p}_{\nu}, h} \Big] \\
& \times \Big[ \mathcal{M}^*_{\ell} \delta(Q - \sqrt{\ul{\mb{p}}_{\nu}^2+m_{\ell}^2}) \bra{\ul{\mb{P}} - \ul{\mb{p}}_{\nu}, \ul{\mb{p}} + \ul{\mb{p}}_{\nu}; \ul{\mb{p}}_{\nu}, \ell} \\
& + \mathcal{M}^*_{h} \delta(Q - \sqrt{\ul{\mb{p}}_{\nu}^2+m_{h}^2}) \bra{\ul{\mb{P}} - \ul{\mb{p}}_{\nu}, \ul{\mb{p}} + \ul{\mb{p}}_{\nu}; \ul{\mb{p}}_{\nu}, h} \Big].
\end{split}
\end{align}
The observable of interest is the probability $P(\mb{P}')$ for the final-state sphere COM momentum $\mb{P}'$, or more accurately the differential distribution $dP(\mb{P}')/d^3\mb{P}'$. We assume that both the internal mode and neutrino are unobservable, and therefore should be traced out. Orthogonality of neutrino flavor eigenstates implies that the mass eigenstates, which are related by a unitary transformation, are similarly orthogonal $\braket{\ell | h} = 0$. Thus the trace over the neutrino state collapses any terms involving mass superpositions.\footnote{Physically this means that an external observer or bath with infinite resolution could projectively measure the neutrino and determine its mass state. Thus if the neutrino is lost to the bath, so is the coherence of information about its mass state.} Taking the trace over both systems and projecting onto a final-state momentum $\bra{\mb{P}'} \cdots \ket{\mb{P}'}$ gives the simple result
\begin{align}
\begin{split}
\label{dgamma}
\frac{d\Gamma}{d^3\mb{P}'} = \int \frac{d^3\mb{p}_{\nu}'}{2\pi} & |\psi(\mb{P}' - \mb{p}_{\nu}')|^2  \Big[ |\mathcal{M}_{\ell}|^2 \delta(Q - \sqrt{\mb{p}_{\nu}'^2+m_{\ell}^2}) \\
&  + |\mathcal{M}_{h}|^2 \delta(Q - \sqrt{\mb{p}_{\nu}'^2+m_{h}^2}) \Big].
\end{split}
\end{align}
Here, the rate $\Gamma = P/T$ means a probability per unit time, where we divided through by an overall time $T$ which appears as a regulator on the square of the energy-conservation delta functions $\delta^2(E) = \delta(E) \times T/2\pi$ (see \cite{Weinberg:1995mt}, chapter 3).

Notice that any effects of the relative degree of freedom have dropped out. This happened because we neglected energy loss to the internal modes (i.e., we approximated the $E_L - E_L' \approx Q$ contribution in the energy conservation delta function, rather than including changes to the kinetic energy of the internal mode), and also because we dropped any dependence on the matrix elements $\mathcal{M}_{\ell,h}$ on the relative mode. For the heavy sterile searches which we focused on here, these effects are negligible. However, going beyond this approximation is important for any measurement of the light neutrino states but would require detailed computations of the solid state interactions. In particular, we have neglected the change of the binding potential $k \to k'$ before and after the decay, caused by the change in net charge of the system in a $\beta$ decay, which appears to have important effects in proposals based on graphene \cite{PhysRevD.105.043502}. We leave these issues to future work in a detailed proposal.

Finally, consider the limit of a very narrow initial state for the sphere motion, $|\psi(\mb{P})|^2 \propto \delta^3(\mb{P})$. This would completely localize the integral over the final state neutrino momentum
\begin{align}
\begin{split}
\frac{d\Gamma}{d^3\mb{P}'} \Big|_{\delta P \to 0} & \propto |\mathcal{M}_{\ell}|^2 \delta(Q - \sqrt{\mb{P}'^2+m_{\ell}^2}) \\
& + |\mathcal{M}_{h}|^2 \delta(Q - \sqrt{\mb{P}'^2+m_{h}^2}),
\end{split}
\end{align}
a perfectly sharp bimodal distribution, with one peak each for the heavy and light neutrino states. In this case, measurement of the final-state sphere momentum could cleanly distinguish the two values. However, if instead $\ket{\psi_{\rm COM}}$ has width $\delta P \gtrsim |m_{\ell} - m_h|$, these two peaks will become blurred out in the integral \eqref{dgamma}. This is the precise sense in which quantum control over the initial COM state for the sphere could be used to solve initial-state broadening effects, as recently highlighted in \cite{Cheipesh:2021fmg,PhysRevD.105.043502,PTOLEMY:2022ldz_theory}.

\end{document}